\documentclass[12pt]{article}
\usepackage{authblk}
\usepackage{geometry}
\usepackage{amsmath}
\usepackage{amssymb}
\usepackage{amsthm}
\usepackage{mathrsfs}
\usepackage{subcaption}
\usepackage{customcommands}
\usepackage{bm}
\usepackage{Nettastyle}
\usepackage{physics} 
\usepackage{tensor}
\usepackage{tcolorbox}
\usepackage{multicol, makecell, longtable} 

\usepackage[normalem]{ulem}

\preprint{MIT-CTP/5646}

\title{Subregion Independence in Gravity}

\author{\AA{}smund Folkestad}
\emailAdd{afolkest@mit.edu}
\affiliation{Center for Theoretical Physics, Massachusetts Institute of Technology, \\Cambridge, MA 02139, USA}

\abstract{
    In gravity, spacelike separated regions can be dependent on each other due to the constraint equations. In this paper, we give a natural definition of subsystem independence and gravitational dressing of perturbations in classical gravity.  We find that extremal surfaces, non-perturbative lumps of matter, and generic trapped surfaces are structures that enable dressing and subregion independence.  This leads to a simple intuitive picture for why extremal surfaces tend to separate independent subsystems. The underlying reason is that localized perturbations on one side of an extremal surface contribute negatively to the mass on the other side, making the gravitational constraints behave as if there exist both negative and positive charges.  Our results support the consistency of islands in massless gravity, shed light on the Python's  lunch, and provide hints on the nature of the split property in perturbatively quantized general relativity. We also prove a theorem bounding the area of certain surfaces in spherically symmetric asymptotically de Sitter spacetimes from above and below in terms of the horizon areas of de Sitter and Nariai. This theorem implies that it is impossible to deform a single static patch without also deforming the opposite patch, provided we assume spherical symmetry and an energy condition.
}

\begin{document}

\maketitle

\section{Introduction}
When are spacelike separated regions in a quantum theory of gravity independent?
This is a central question that holography forces us to confront. On the one
hand, the constraint equations of GR tells us that some gravitational
subregions depend on each other, since generically one cannot specify initial
data on two regions independently. On the other hand, the 
theory of entanglement wedges and subregion-subregion duality
\cite{RyuTak06,RyuTak06-2,HubRan07,CzeKar12,Wal12,HeaHub14,EngWal14,AlmDon14,HaPPY,JafLew15,Har16,DonHar16,CotHay17}
in AdS/CFT demonstrates that independent regions do exist -- at least
perturbatively in the $G_N \rightarrow 0$ limit. Thus, some
spatially separated regions are mutually dependent, while others are not. 

For perturbative canonical quantization of GR around flat space, any compact region in the
bulk in some sense depends on a region surrounding spatial infinity, since
every compactly supported operator must be dressed to infinity when coupled to
gravity \cite{DonGid16}, due to diffeomorphism invariance. No
excitation can be turned on in the center of the spacetime without disturbing
the gravitational field at infinity. An even stronger notion of dependence on
infinity has been shown for perturbations of pure AdS: to first order in perturbation theory, two Wheeler-deWitt (WdW) functionals
that agree on observables in a time-band near the boundary must be equal
\cite{ChoGod21}. It might be tempting to think that perturbatively quantized GR
itself is holographic, in the sense that perturbative data near the conformal boundary
can be used to access perturbative data in the deep bulk, without relying on
non-perturbative corrections in $G_N^{-1}$.\footnote{See \cite{Bal06,Mar06,Mar08b,
Mar13} for the development of the idea that a version of holography is implied by
the fact that the Hamiltonian is boundary term, and also for discussion of the importance of
non-perturbative corrections. See also \cite{ChoPap21,LadPra21} for recent work on these ideas. } But what about other
backgrounds? While gravity contains no truly local operators, once the
background has sufficiently rich structure, there exist quasilocal operators \cite{Kom58,BerKom60,GidMar05,Tam11,Mar15,Kha15,GoeHoe22} that are
dressed features other than conformal infinity. For such backgrounds,
one can construct operators that commute with operators near the
boundary to all orders in perturbation theory \cite{Mar15, BahBel22,BahBel23}. 
This shows that we must be cautious about extrapolating lessons about subregion
dependence learned from working perturbatively around featureless spacetimes
like Minkowski or AdS. At the classical level,
these satisfy rigidity theorems
\cite{SchYau81,Wit81,Wan01,ChrHer01} that
strongly constrain the nature of perturbations around these spacetimes. 

\begin{figure}
\centering
\includegraphics[width=0.5\textwidth]{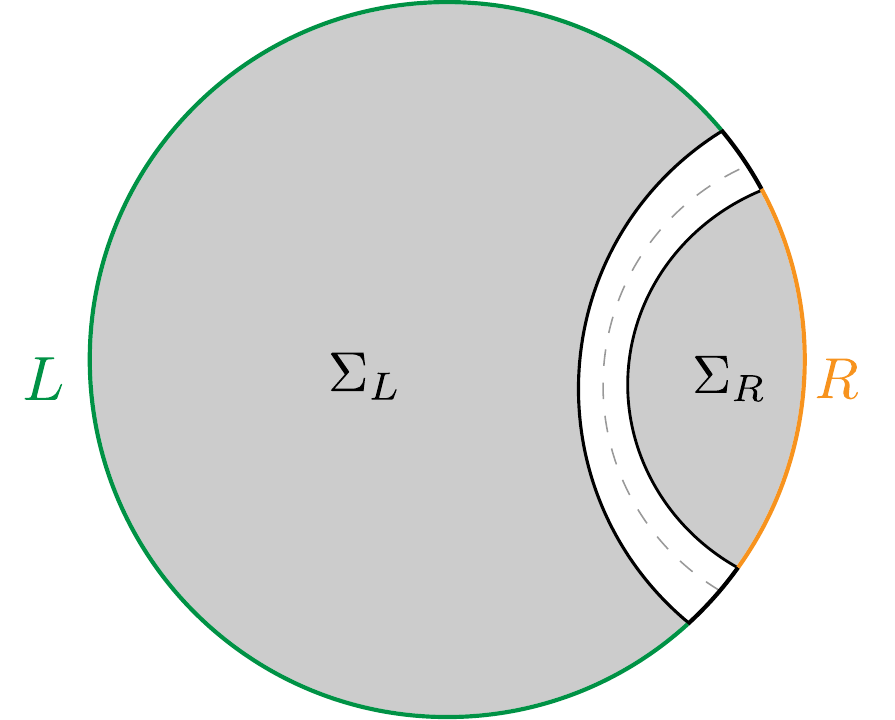}
    \caption{Illustration of an AdS Cauchy slice $\Sigma$ containing Cauchy
    slices $\Sigma_L, \Sigma_R$ for the entanglement wedges of the boundary
    regions $L, R$. Taking the classical limit of AdS/CFT, we expect that we can
    make perturbations of the classical initial data in $\Sigma_L$ and
    $\Sigma_R$ independently.
    }
\label{fig:adsintro}
\end{figure}

In this paper, we will show that the question of subregion independence is
surprisingly interesting and illuminating already at the classical level. We
will demonstrate that the nature of the background strongly influences how and when
regions become independent. This will lead to a simple physical picture where
classical extremal surfaces play an important role in ensuring subregion
independence.

To approach the precise question we will ask in this paper,
consider two spatial boundary subregions $L, R$ on the conformal boundary of
an asymptotically AdS (AAdS) spacetime. Assume that
these are spacelike separated by a small but finite gap -- see
Fig.~\ref{fig:adsintro}. 
By the so-called split property of local QFT \cite{Roo70,Buc74,DopLon84,BucAnt87,Few16}, we then have a notion of independence 
for the CFT states on $L, R$. Namely, for any two independent choices of states
$\omega_L$
and $\omega_R$ on these regions, there exists a pure state $\ket{\psi}$ on the
full Hilbert space that agrees with $\omega_L, \omega_R$ on $L,
R$:\footnote{Strictly speaking, the split property guarantees a state on $R \cup
L$ with this property. But we expect that this state can be purified on the
full Hilbert space.}  
\begin{equation}
\begin{aligned}
\bra{\psi}\mathcal{O}_{L}\mathcal{O}_{R}\ket{\psi} =
    \left<\mathcal{O}_{L}\right>_{\omega_L}\left<\mathcal{O}_{R}\right>_{\omega_R},
\end{aligned}
\end{equation}
where $\mathcal{O}_L, \mathcal{O}_R$ are operators in the algebra of (the domain of
dependence of) $L, R$, respectively.
Assuming we are in a holographic CFT, 
$\omega_L, \omega_R$ encodes data about the entanglement wedges of $L, R$. 
Now, let $\Sigma_L, \Sigma_R$ be Cauchy slices of the entanglement wedges of $L,
R$,
contained in a full Cauchy slice $\Sigma$ shown in Fig.~\ref{fig:adsintro}. In the strictly
classical limit of AdS/CFT, we expect that $\Sigma_L,
\Sigma_R$ inherit some notion of independence from the QFT split property. A
natural expectation would be the following: for any two small perturbations of
initial data in $\Sigma_L, \Sigma_R$, there exists some complete set of GR
initial data on all of $\Sigma$ that agrees with our perturbed data on
$\Sigma_L, \Sigma_R$. Furthermore, this complete initial data is perturbatively close to
the initial data we started with. In more pedestrian language: we can wiggle the initial
data in the two subregions independently, without making drastic
non-perturbatively large changes to the full spacetime.\footnote{In AdS/CFT
language, we do not have to leave the classical limit of our code subspace where the entanglement
wedges were defined.} However, there is no one-liner argument in GR
showing that this is true, since the constraint equations are elliptic -- they
have no lightcones. Generically, modifications of matter (or pure gravity
degrees of freedom) in some compact region
$A$ source gravitational fields spacelike to $A$, so how do we know that there
are no perturbations of initial data in $\Sigma_L$ that requires us to change
fields in $\Sigma_R$? We might be tempted to mumble words about dressing
observables to $R$ and $L$, but this is not quite satisfactory. First, we should be able to ask this
question at the level of solutions of the classical constraint equations of GR. Second, and more
importantly, this leaves no room for the extremal surface playing any kind of
special role -- this explanation would lead us to conclude that any two disjoint regions
$\Sigma_1, \Sigma_2 \subset \Sigma$ connected to the asymptotic boundary and separated by a gap
are independent. It is not clear that this is true.

\begin{figure}
\centering
\includegraphics[width=0.5\textwidth]{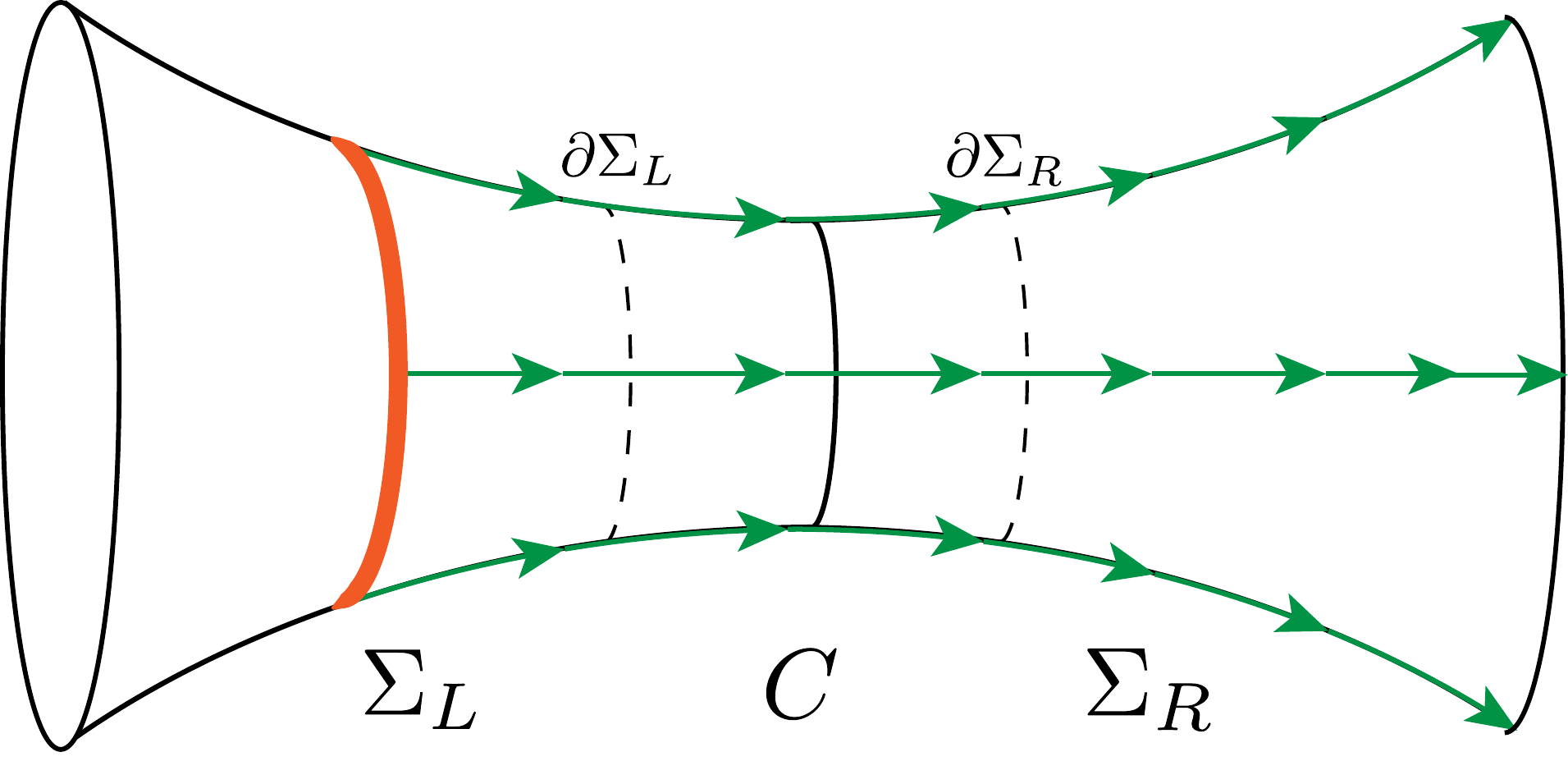}
    \caption{Initial data for electromagnetism coupled to matter on a fixed AdS-Schwarzschild
    background, on a canonical $t=0$ slice. The regions $\Sigma_L, \Sigma_R$
    are not independent when electromagnetism is coupled to charge of a fixed
    sign. If we pick initial data on $\Sigma_L$ corresponding to adding a
    charged shell of matter, and with the $E^i=0$ at left infinity, the Gauss
    law imposes that $E^i \neq 0$ at right infinity.
    }
\label{fig:introfig}
\end{figure}

To illustrate the problem more clearly, let us for a brief moment consider a
different example where we model gravity as
electromagnetism coupled to matter with only positive charge density, living on
a fixed background geometry. In this
case, Gauss' law is a stand-in for the Hamiltonian constraint of GR, the electric
field $E^i$ for the gravitational field, and the charge density $\rho$ for the
energy density. Consider now this theory on a fixed AdS-Schwarzschild background,
and let $\Sigma_L, \Sigma_R$ be the left and right sides of the bifurcation
surface on a canonical $t=0$ slice. We let $\Sigma_L, \Sigma_R$ terminate
slightly to the left and right of the bifurcation surface, so that there is a small
open region $C$ between these slices that contains the bifurcation
surface.\footnote{This case has a qualitative difference from the previous
example. $\Sigma_L$ and $\Sigma_R$ are slightly smaller than Cauchy slices of the
 entanglement wedges of the left and right boundaries, $R, L$. This is
because $\omega_L, \omega_R$ now cannot be chosen completely
independently. If their entanglement entropies disagree, there is no pure state
$\ket{\psi}$ on the joint Hilbert space that reduces to $\omega_R, \omega_L$. So we
leave a small number of degrees of freedom flexible, to avoid this obstruction.
} 
See Fig.~\ref{fig:introfig}. A
perturbation of the trivial $E^i=\rho=0$ initial data on $\Sigma_{L}$ is the
following: add a spherically symmetric shell of charged matter and choose that
the electric field is unchanged near left infinity. This is compatible with
Gauss' law on $\Sigma_L$. However, now Gauss' law requires
the new electric field lines sourced by the shell to travel towards the right.
Because our theory contains no negative charges, there is nothing we can do in $C$ to absorb
the field lines before they reach $\Sigma_R$. They will travel all the way to right infinity.
See Fig.~\ref{fig:introfig}.
The initial data on $\Sigma_L$ and $\Sigma_R$ cannot be chosen independently. 

Now let us turn to gravity proper. If the Hamiltonian constraint 
behaved like Gauss' law coupled to matter with only positive
charge densities, this would not bode well for the independence of initial data
perturbations in $\Sigma_R$ and
$\Sigma_L$ in gravity. However, except in the special case of perturbations around pure AdS or Minkowski, 
Gauss' law coupled to purely positive charge densities is a poor
analogy for the Hamiltonian constraint.  
Crucially, the total gravitational mass is not
just an integral of the local energy density weighted by something
positive. 
As we will explain, extremal surfaces play a very practical role in
ensuring that complementary entanglement wedges are independent (after you cut
out a small splitting region). Specifically, we are going
to show that if you live in $\Sigma_R$, positive energy density added to
the other side of the bifurcation surface, from your perspective, acts like a negative energy
density! Furthermore, this effect is present for general extremal surfaces. 
This means that positive energy densities can be used to
terminate gravitational fields sourced by positive energy
densities on the opposite sides of an extremal surface.\footnote{Strictly
speaking, we
are able to show this in all dimensions only with spherical symmetry, while in four
spacetime dimensions we are able to remove this assumption.}  For compact extremal
surfaces, this effect cannot be
understood by studying the Hamiltonian constraints perturbatively around AdS or
flat space, or by relying on analogies to Gauss' law. It relies on the
non-linear nature of gravity, which causes structures of the background to
significantly affect when subregions allow independent perturbations of initial
data. We argue that this effect relieves a tension
that was argued to exist between islands and massless gravity in
\cite{GenKar21} (see also \cite{BahBel22,BahBel23} for arguments that islands
and massless gravity are consistent). It also ensures the independence of $\Sigma_L$ and $\Sigma_R$
in the example considered above.

Next, we will prove or argue, depending on the setup, that trapped
regions of spacetime always have a kind of indeterminate energetic behavior.
From the perspective of the mass in any asymptotic region, there are always some
modes in a trapped region that, when turned on, increase the mass, and others
that reduce it.\footnote{This has of course long been understood for stationary
black holes, where the Killing vector that defines the asymptotic
energy flips signature in the bulk, but we will make it clear more
generally that trapped regions always have this behavior.} Together, these
enable perturbations of spatial compact support. This gives subregions
of a trapped region a larger degree of independence from nearby regions, and we
will use it to argue that the islands found in the evaporating black holes of \cite{AlmEng19,Pen19} easily can host a large number of localized
excitations that never affect the black hole exterior. Along the way we will
also show the following fact: an object's contribution to
the ADM mass approaches zero as the object localizes on an extremal surface (see
Secs.~\ref{sec:lorentzian}, \ref{sec:IMC}, for precise statements). So around an extremal surface,
we can add arbitrarily high energy objects (as measured locally) with
no or arbitrarily small change
in the ADM mass.\footnote{This plays nicely with the ideas of \cite{Har15},
which propose that Wilson lines threading a wormhole really, in the bulk UV, are
split into factorizing operators by heavy charges residing near the bifurcation surface. 
This ensures factorization. The fact that general classical extremal surfaces can host heavy
objects without large cost in energy supports the viability of this idea in a large class
of backgrounds.
}

In the following, we first give a precise definition of subregion independence in
GR in Sec.~\ref{sec:independence} and argue why it is a useful definition.  We
explore our definition in a series of
illuminating examples in Sec~\ref{sec:examples}. We find that regions separated
by extremal surfaces ``shields'' of matter tend to
classically independent. Sometimes trapped surfaces can play a similar role. 
Of course, we do not expect that this independence generally persists (or can be
defined)  in full non-perturbative quantum gravity, but it does appear plausible that this
independence could be inherited by a perturbative quantization of GR around a
fixed background.

Then, in Sec.~\ref{sec:general}, we conduct a more
general analysis. For spacetimes with spherical symmetry, we prove that the
behaviors we found in the examples are general: gravity truly behaves
as if it has negative and positive energies when extremal or trapped surfaces are
present. In four-dimensional spacetime, we are also able to remove the spherical
symmetry assumption, giving strong arguments at the physics level of rigor. 
We do expect that the physical effects we identify to survive in other
dimensions, although we do not know what are the right tools to show this in
spacetimes without symmetries. 

Then, guided by findings when studying independence of subregions of de Sitter in
Sec.~\ref{sec:examples}, in Sec.~\ref{sec:dS} we prove a new theorem upper and lower bounded the area of a family of
surfaces that include certain extremal and trapped surfaces in asymptotically dS
spacetimes. A corollary of the theorem states that under spherical symmetry and given an
energy condition, adding matter in one static patch necessitates adding matter
in the opposite patch. Furthermore, adding matter always reduces the area of a certain
extremal/trapped surface. This result holds at the full non-linear level.

Finally, in Sec.~\ref{sec:discussion} we discuss implications of our findings
for semiclassical gravity, quantum extremal surfaces, islands in massless
gravity, localized observables, and the Python's lunch.

\section{Subregion independence}\label{sec:independence}
\subsection{The constraint equations}
To understand subsystem independence, we need to deal with the
gravitational constraint equations. Let $\Sigma$ be a manifold, possibly with
boundary, and let
\begin{equation}
\begin{aligned}
S \equiv (h_{ab}, K_{ab}, \Phi)
\end{aligned}
\end{equation}
be a regular initial dataset for the Einstein equations.  Here $h_{ab}$ is a Riemannian
metric on $\Sigma$, $K_{ab}$ a symmetric tensor, and $\Phi$ a label that
collectively denotes initial data for all of the matter fields. For example,
when the matter is  a scalar field, $\Phi$ consists of the value of the scalar and its
time-derivative on $\Sigma$. For any choice of $S$, 
assuming the matter is reasonably well behaved, there is a unique (up to
diffeomorphism) globally 
hyperbolic spacetime $(M, g)$ corresponding to the maximal evolution of
$(\Sigma, S)$
\cite{Cho52,ChoGer69,HawEll,Cho09}. The pair 
$(\Sigma, S)$ represents a moment of time in $(M, g)$, with $\Sigma$ being an embedded
spatial slice, and with $h_{ab}, K_{ab}$ being the induced metric
and extrinsic curvature of $\Sigma$, respectively.
Note that we never consider initial data such that $\Sigma$ has a kink or ends
in a singularity. If we extract an initial data slice $(\Sigma, S)$  from an existing
spacetime $(N, g)$, unless otherwise specified, we always assume it is globally (AdS-)hyperbolic, with $\Sigma$ being a Cauchy slice.\footnote{For AAdS
spacetimes, which are technically not globally hyperbolic,
we assume the AdS notion of global hyperbolicity \cite{Wal12}.}

Due to diffeomorphism invariance, initial data is not freely specifiable. 
Instead, it has to satisfy a set of constraint equations. In the
case of Einstein gravity minimally coupled to matter, the constraint
equations read (in units of $8\pi G_N=1$)
\begin{align}
    \mathcal{R} - K_{ab}K^{ab} + K^2 - 2\Lambda = 2 \mathcal{E},
    \label{eq:Econs}\\
    \mathcal{D}^b K_{ba} - \mathcal{D}_{a} K =
    \mathcal{J}_a,\label{eq:Jcons}
\end{align}
where $\mathcal{R}$ is the Ricci scalar of $h_{ab}$, $\mathcal{D}_a$ the 
$h_{ab}$-compatible connection on $\Sigma$, $K=K\indices{^a_a}$, and $\Lambda$ the
cosmological constant. $\mathcal{E}$
and $\mathcal{J}_a$ are the matter energy and momentum densities,
respectively. From the spacetime perspective, if $n^a$ is the future unit normal to
$\Sigma$, $P_{ab}$ the projector onto the tangent space of $\Sigma$, and $T_{ab}$ the matter stress tensor, then $\mathcal{E}=T_{ab}n^a n^b$
and $\mathcal{J}_a = T_{cb}n^{b}P\indices{^{c}_a}$. Equations \eqref{eq:Econs}
and \eqref{eq:Jcons} are known as the Hamiltonian and momentum constraints,
respectively. Unless otherwise stated, we will assume the weak energy condition (WEC), which
says that the local energy density is positive:
\begin{equation}\label{eq:WEC}
\begin{aligned}
\mathcal{E} \geq 0.
\end{aligned}
\end{equation}
As will become clear soon, we are in an unusual situation where we are assuming
the WEC to make our lives harder, not easier. The positivity of
local energy densities is a
significant obstruction to achieving independence of subregions in gravity, so
if two regions are independent despite the WEC, then we expect them to be
independent also when the WEC is violated.\footnote{Modulo other complications that might
arise if WEC breaking comes from considering quantum effects.}

As alluded to in the introduction, the central property of the constraint equations for us is that they are not
hyperbolic equations. There are no lightcones. A change in the matter fields
with support on some region $A\subset \Sigma$ might demand a change in
the gravitational or gauge fields in a different spacelike separated region
$B\subset \Sigma$.
This type of behavior is of course already present in the Maxwell equations.
For example, at $t=0$ in Minkowski spacetime, it is impossible to add an
electron without also adding an electric field which either reaches infinity, or
alternatively, also adding a positron that the field can terminate on.
From the QED perspective, this is because there are no
gauge-invariant operators
that create a single electron at a point $x$. Instead one needs to dress the operator to
restore gauge invariance, for example by adding a Wilson line that reaches out
to infinity:
\begin{equation}
\begin{aligned}
    e^{i\int_{x}^{\infty} A} \psi(x),
\end{aligned}
\end{equation}
where $A, \psi$ are the gauge and Dirac fields, respectively.  This adds
precisely the required electric field to satisfy the constraints. However, there
is no need study the quantum theory to see the need for dressing -- the effect
is already there in the classical Gauss' law.  This suggests that a careful
understanding of the constraints, even at the classical level, can give
hints about what gauge invariant operators, and thus what algebras of
observables, can exist.  This is an additional motivation for this study.  We
will see the Einstein constraint equations yield
behavior significantly richer than Gauss' law.

\subsection{Subregion independence and dressing}
To define subsystem independence, we need to look at perturbations to the
constraints. We will use $\delta S \equiv (\delta h_{ab}, \delta K_{ab}, \delta
\Phi)$ to denote a regular perturbative solution of the constraint equations around some
initial dataset $S$ ($\Sigma$ is often left implicit). The small perturbation parameter is the amplitude
of the perturbation. Unless otherwise specified, we assume $\delta S$ is a
formal series solution to all orders in perturbation theory in this amplitude,
and not just a linearized solution. We always choose the lowest order terms in the perturbation so
that the leading order correction to the stress tensor and the metric is at the
same order. Note that some perturbations do not correspond to a physical
change of the spacetime, but instead to a diffeomorphism, either changing
coordinates within $\Sigma$, or moving us to a different slice nearby $\Sigma$ in the same
spacetime. We allow these.

Next, consider a subregion $A\subset \Sigma$. We will use the notation $\delta
S|_{A}$ to refer to a perturbative constraint solution restricted to $A$,
and we will refer to a perturbative constraint solution $\delta S$ specified on all of
$\Sigma$ as an \textit{extension} of $\delta S|_A$ if it restricts to $\delta
S|_A$ on $A$. If an extension exists, it will generally not be unique.
If we specify a perturbation of the constraint on two disjoint
subregions $A,B$, we take $\delta S|_A \cup \delta S|_{B}$ to mean the obvious
concatenated perturbation
on $A \cup B$. Finally, since we have in mind working in some particular theory, when we talk about
perturbations, we only ever talk about perturbations that respect the boundary
conditions imposed on the boundaries of $\Sigma$ by our theory -- either at finite
boundaries or at spatial infinity. 

We are now ready to define when $A$ and $B$ are independent:
\begin{defn}[Independence]\label{def:main}
    Let $(\Sigma, S)$ be an initial dataset, and let $A, B\subset \Sigma$ be
    two disjoint subregions. We say that $A$ and $B$ are independent if for any
    two perturbations $\delta S|_A$ and $\delta S|_B$, there exists a
    perturbative extension
    $\delta S$ of $\delta S|_A \cup \delta S|_B$ to all of $\Sigma$.
\end{defn}
Definition \ref{def:main} simply says that it is possible to wiggle
the initial data in $A, B$ completely independently, provided these wiggles are
perturbatively small.\footnote{Rather than perturbative deformations, we could
alternatively work with finite but arbitrarily small deformations with
respect a $C^k$ or Sobolev norm defined by the background metric, as is common
in mathematical studies of the constraints. See for example \cite{HawEll,Cho09} for a
discussion of such norms. We expect that the physical effects we discuss in
this paper are robust to taking this approach instead.}
If $A$ and $B$ are not independent, we say that they
are dependent. If $A$ and $B$ are independent, it is natural to say that
classical gravity satisfies the split property across $C\equiv \Sigma-A-B$. 
This means that for any (perturbative) choice of the classical state on $A$ and $B$, there exist a classical state on all of $\Sigma$ that
agrees with the chosen states on $A$ and $B$.
Note that it is important that we
work within a fixed theory, which comes with a set of allowed matter fields and
boundary conditions. We might think that we can use gluing constructions
supported by delta functions shocks to make more or less anything
independent, but this is not so. Gluings will generically
produce distributional stress tensors that do not satisfy any energy conditions.
Even if they do, there is no guarantee that regularizations of these shocks can
be achieved by matter in our theory. So we might as well stick with genuinely regular initial
data. 

Next, note that since pure diffeomorphisms are allowed, perturbing the
boundaries of the regions $A, B$ count as perturbing the initial data in $A$ and
$B$.  To see this, consider the data $S$ and assume that the regions $A$, $B$ are fixed in
some gauge. Assume now that a
diffeomorphism $\psi: \Sigma \rightarrow \Sigma$ maps the boundaries of $A, B$
to new locations. An initial dataset in which the boundaries have moved, but
with the spacetime and the slice $\Sigma$ left fixed, is $(\Sigma, \psi[A], \psi[B],
S)$. By diffeomorphism
invariance, this is equivalent to $(\Sigma, A, B, \psi_*[S])$, where $\psi_*$ is
the pull-back.\footnote{Note that if $\psi$ is a ``large diffeomorphism'' (not
large in
the perturbative sense, but in the sense that it has non-trivial action on the
boundary of spacetime), then we
have both changed the full physical state, in addition to
moving $A$ and $B$.}
This logic can
easily be generalized to a diffeomorphism perturbing $A, B$ out of the slice.
Thus, the regions $A$ and $B$ are not completely fixed -- they are fuzzy at the
perturbative scale. Alternatively said, the perturbative edges modes \cite{DonFre16} of $A$ and $B$ can be
activated when testing for independence. Before we apply a physical
perturbation in $A$ that changes the spacetime, we are allowed to deform the
boundaries of $A$ by a perturbative amount first -- or the other way around. 

Definition \ref{def:main} can be applied to
classical non-gravitational theories as well. For theories without constraints,
such as a scalar
field $\phi$, all regions $A, B$ separated by an open set $C$ are independent. We
can always match the initial data $\phi, \dot{\phi}$ in $A, B$ smoothly across
$C$, since they are completely unconstrained there. However, the \textit{splitting
region} $C$ was important. Even in free scalar theory, two regions with
intersecting closures are dependent. Next, even in the
presence of a constraint like Gauss' law of electromagnetism, on Minkowski space
 all spacelike separated
 subregions are completely independent, provided we
 couple to charged matter with both signs, as explained in the introduction.
 If we add a positive
charge density in $A$, we just need to make sure to compensate with a negative charge density
in $C$ to screen any new multipole
moments incurred in $B$.\footnote{
In vacuum electromagnetism
 on Minkowski, we also have independence of subregions
 \cite{BeiChr17}.
    However, on non-trivial manifolds, the question of independence is more
    interesting. For example, as pointed out in \cite{HarOog18}, vacuum
    electromagnetism on a spatial torus will not have subregion independence.
    For example, on $T^2$, this is because the value of the electric charge on two
    homologous non-contractible $S^{1}$'s must match in different regions.
}

In gravity things are more interesting. Because of the positivity of local energy densities,
it is generally
harder to screen perturbations, and the question of independence is non-trivial.
As an example, consider $\Sigma$ to be a canonical
$t=0$ slice of Minkowski or AdS. Let $A\subset \Sigma$ be a compact region and
$B \subset \Sigma$ an
infinite annulus containing spatial infinity. By the positive mass
theorem (PMT) \cite{Wit81,SchYau82,Wan01,ChrHer01,AndCai08}, any perturbation in $A$ which is not a pure
diffeomorphism must increase the spacetime's mass. Since the ADM Hamiltonian is
a boundary term localized to $B$, this implies that the gravitational data in
$B$ must be perturbed as well. Thus $A$ and $B$ are dependent. Pure AdS and Minkowski support
no perturbations of compact support, except for pure gauge.\footnote{It is
crucial to not just work to linear order in perturbations, since compact
perturbations of Minkowski exist at linear order \cite{BeiChr17}.} 
From the perspective of perturbative quantum gravity around flat space and AdS,
this can be seen as the classical avatar of the statement that
operators must be dressed to infinity \cite{DonGid16}, for example with a gravitational Wilson
line (see for example \cite{GidKin18,GidWei19,HarWu21,GidPer22}). Effectively, there is no negative energy density to
terminate the Wilson line on.

Going away from pure AdS or Minkowski, since matter has positive energy
density,
it naively seems like we are always forced to dress a perturbation to asymptotic
infinity, in turn causing every compact subregion to depend on an annulus around
spatial infinity. However, this conclusion is much too quick. When certain
geometric features are present, such as extremal and trapped surfaces or lumps
of matter, localized dressings become possible.

To streamline the discussion going forward, it is useful to give a precise
notion of classical dressing of a perturbation (as opposed to an observable):
\begin{defn}[Dressing] 
    Let $(S,\Sigma)$ be an initial dataset. Let $A$ be a subregion and $\delta S|_A$
    a perturbation of the constraints on $A$. If $\delta S$ is an extension of
    $\delta S|_A$, we say that $\delta S$ is a dressing of $\delta S|_A$
    to the region $\text{supp}(\delta S)$.
\end{defn}
We should emphasize that given some $\delta S|_A$, the choice of gravitational
dressing is highly non-unique. 
The independence of $A$ and $B$ 
implies that for every perturbation of $A$, there exists some choice of
dressing that does not intersect $B$. Furthermore, after dressing any
perturbation on $A$ to $C$, we still have the ability to dress any perturbation in $B$
to $C$ as well. The statement of independence is not that a certain
perturbation in $A$ could not be dressed to $B$, if so desired.

Next, let us address two questions that naturally arise
from our definition. What about regions in
spacetime that are not conveniently described as lying on the same slice? 
Why only perturbative deformations?

\begin{figure}
\centering
\includegraphics[width=0.9\textwidth]{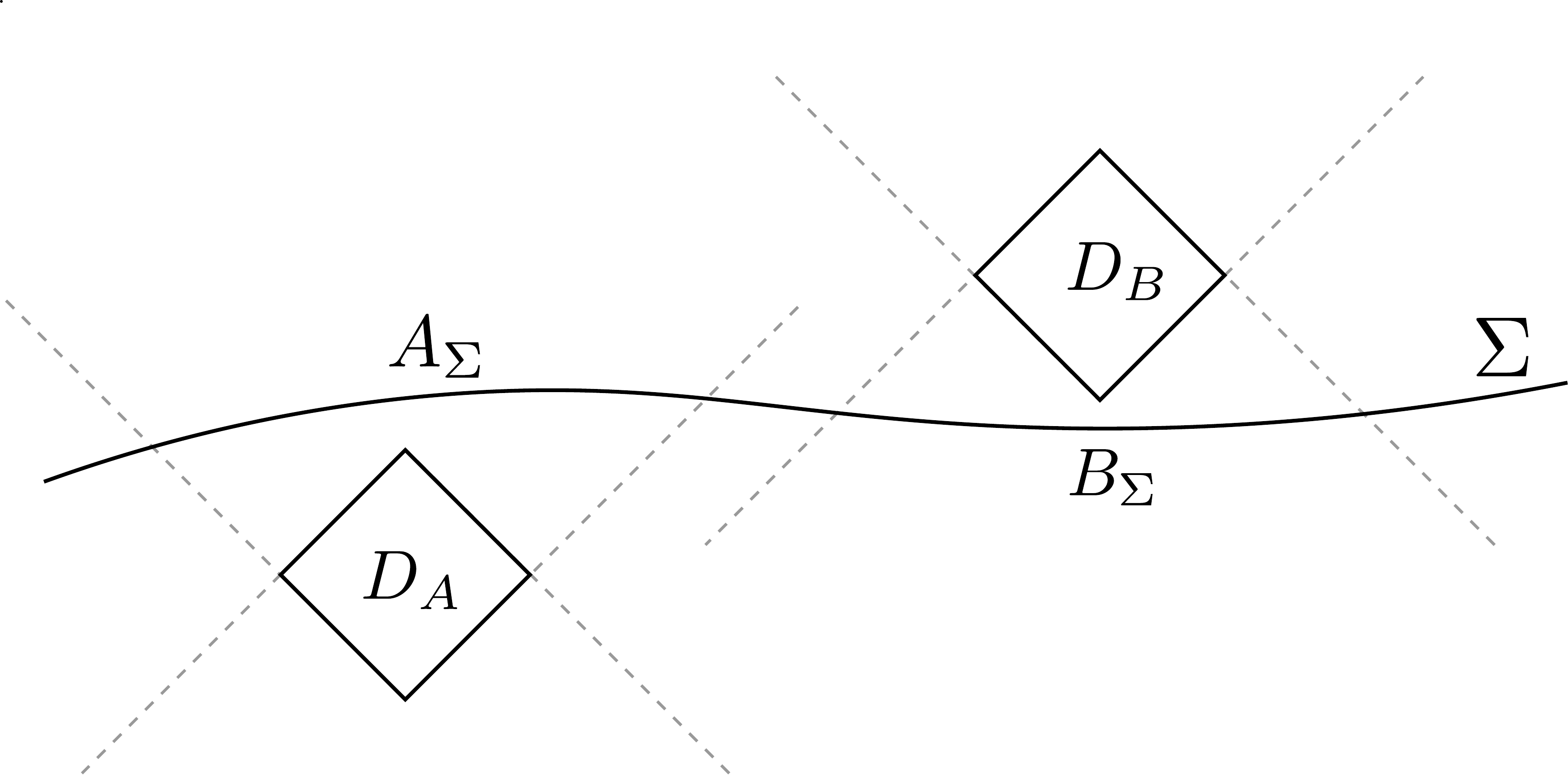}
    \caption{$D_A$ and $D_B$ are independent if there exist a slice $\Sigma$ such
    that $A_{\Sigma}$ and $B_{\Sigma}$ are independent.}
\label{fig:diamond}
\end{figure}

First, in globally (AdS-)hyperbolic spacetimes,\footnote{We can also apply this definition within a causal
diamond of an AdS spacetime without reflecting boundary conditions.} it is straightforward to extend our notion of independence to two causal diamonds $D_A$
and $D_B$. First, if $D_A$ and $D_B$ overlap or are timelike separated, they
should clearly not be considered independent, so assume that they are
spacelike separated. Let $\Sigma$ be any (AdS-)Cauchy slice $\Sigma$, and define
\begin{equation}
\begin{aligned}
    A_{\Sigma} &= \Sigma \cap (J^+[D_A] \cup J^-[D_A]), \\
    B_{\Sigma} &= \Sigma \cap (J^+[D_B] \cup J^-[D_B]),
\end{aligned}
\end{equation}
where $J^{+}[X]$ ($J^{-}[X]$) is the usual causal future (past) of a given
set $X$. We say that $D_A$ and $D_B$ are independent if there exists a Cauchy slice
$\Sigma$ such that $B_{\Sigma}$ and $A_{\Sigma}$ are independent, in the
sense already defined. See Fig.~\ref{fig:diamond}. This is natural, since by causality and global
hyperbolicity, the gravitational initial data on
$A_{\Sigma}, B_{\Sigma}$ contains all information about $D_A, D_B$.  Thus, any
perturbation in $D_A$ and $D_B$ can, through the evolution equations, be
translated into a corresponding perturbation on $A_{\Sigma}$ and $B_{\Sigma}$.

Second, the reason that we only consider perturbative deformations is that if we
allow large deformations of the initial data, there is no sense in which the
regions are fixed. If we
could modify the data in $A, B$ arbitrarily much, we could deform it to any
data that is compatible with the topologies of $A, B$, or use a diffeomorphism to
move the boundaries of $A, B$ arbitrarily much. But then we would in the end
only be asking the following question: does there exist interpolating initial
data between every possible choice of initial data on two regions with the
given topologies of $A$ and $B$.  But why do we even care to keep
topologies fixed at this point? Thus, going beyond the perturbative level, we
quickly lose all
connection to our original question. This problem is reflective of the deeper underlying
fact that, on the global phase space of GR, there is probably no useful
 diffeomorphism invariant notion of a subregion. However, when we restrict to
perturbative deformations, we let the regions and the states
become fuzzy, but only at the scale of the perturbative parameter. In language
used in AdS/CFT contexts, asking about independence of two spacetime
regions seems to be a  natural question only in a code subspace built of
perturbations around a single background (and, at the
quantum level, in the small$-G_N$ limit).

It is worth making a few more comments on the non-perturbative case.
However, this is a detour, and the reader can freely skip the next section without any loss of
coherence in the rest of the paper.

\subsection*{Gluing constructions and independence}
\begin{figure}
\centering
\includegraphics[width=0.8\textwidth]{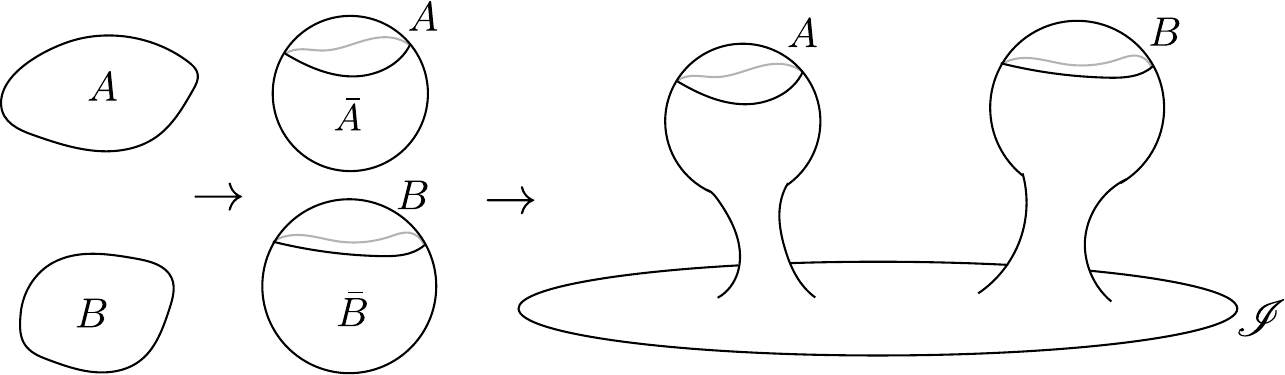}
    \caption{Using the gluing construction of \cite{ChrPio04} to embed two
    compact datasets in a complete dataset with a pre-specified conformal infinity $\mathscr{I}$.}
\label{fig:gluing}
\end{figure}

Consider making small deformations of initial data in $A, B \subset
\Sigma$, but
not requiring that the change outside $A \cup B$ is small -- i.e. we allow
 jumping to a point in the GR phase space that is not near our original point. Can we
then find an initial dataset interpolating between $A$ and $B$, and which still
has the requisite number of conformal boundaries? While
this question has little relevance to understanding
perturbative quantization around a background, it is nevertheless interesting
to see what happens in this case. In vacuum gravity, it can be partially answered, 
thanks to the gluing result of \cite{ChrPio04}. This result says that in
vacuum gravity, if $\Sigma$ is a smooth initial dataset
and $\Omega_1, \Omega_2 \subset \Sigma$ are two open sets whose domains of
dependence have no Killing
vectors, then we can cut out geodesic balls in
both $\Omega_1, \Omega_2$ and glue in a handle $[0, 1] \times S^{d-1}$ to
connect $\Omega_1, \Omega_2$ with a wormhole. Furthermore, this can be achieved so that the
final initial data is smooth and unchanged outside
$\Omega_1 \cup \Omega_2$. Using this, if $A, B \subset \Sigma$
 are compact  and sufficiently generic, then we can 
embed them in a common initial
dataset through the following procedure: adjoin $A$ and $B$ to some compact
initial datasets 
$\bar{A}$ and $\bar{B}$ such that $\bar{A}\cup A$ and $\bar{B} \cup B$ are
complete compact initial datasets, with $\bar{A}, \bar{B}$ both having at least one
neighbourhood whose domain of dependence has no Killing vectors. Then take
another sufficiently generic initial dataset with the required number of
conformal boundaries, and use the gluing result to attach two wormholes to it -- one
connecting to $\bar{A}$ and one to $\bar{B}$. See Fig.~\ref{fig:gluing}. One can see that a similar
procedures works if $A$ and/or $B$ merely have compact boundaries, so that $A$
and $B$ might contain complete connected components of conformal infinity. However, if
$\partial A$ and $\partial B$ themselves are anchored to conformal boundaries, while we
again can find a common initial dataset containing $A$ and $B$ using gluings, the
complete set of initial data might have too many conformal boundaries, since the
naive type of procedure described above will not connect any of the 
 subregions of the conformal boundary that might be present in $A$ and $B$.

\section{How background structures enable independence}\label{sec:examples}
In this and the following section, we will show how extremal surfaces, matter,
and generic trapped surfaces can be used to dress perturbations, avoiding dressing to
asymptotic regions. They are background structures that enable subregion
independence. We will first study a set of illustrative examples in this section, where
we restrict to particular backgrounds with spherical symmetry.  Then, in
Sec.~\ref{sec:general} we will give a more general discussion, relaxing many of
the simplifications made in our examples.

\subsection{Setup}\label{sec:setup}
Our example will be Einstein gravity in $d+1$ spacetime dimensions coupled to a
massless scalar, with action
\begin{equation}
\begin{aligned}
    I = \int \dd^{d+1}x\sqrt{-g}\left[R+\frac{ d(d-1) }{ L^2 } - \nabla_a \phi
    \nabla^a \phi\right].
\end{aligned}
\end{equation}
For positive cosmological constant, $L$ is imaginary, and we set $L=iL_{\rm dS}$.

\begin{figure}
\centering
\includegraphics[width=0.8\textwidth]{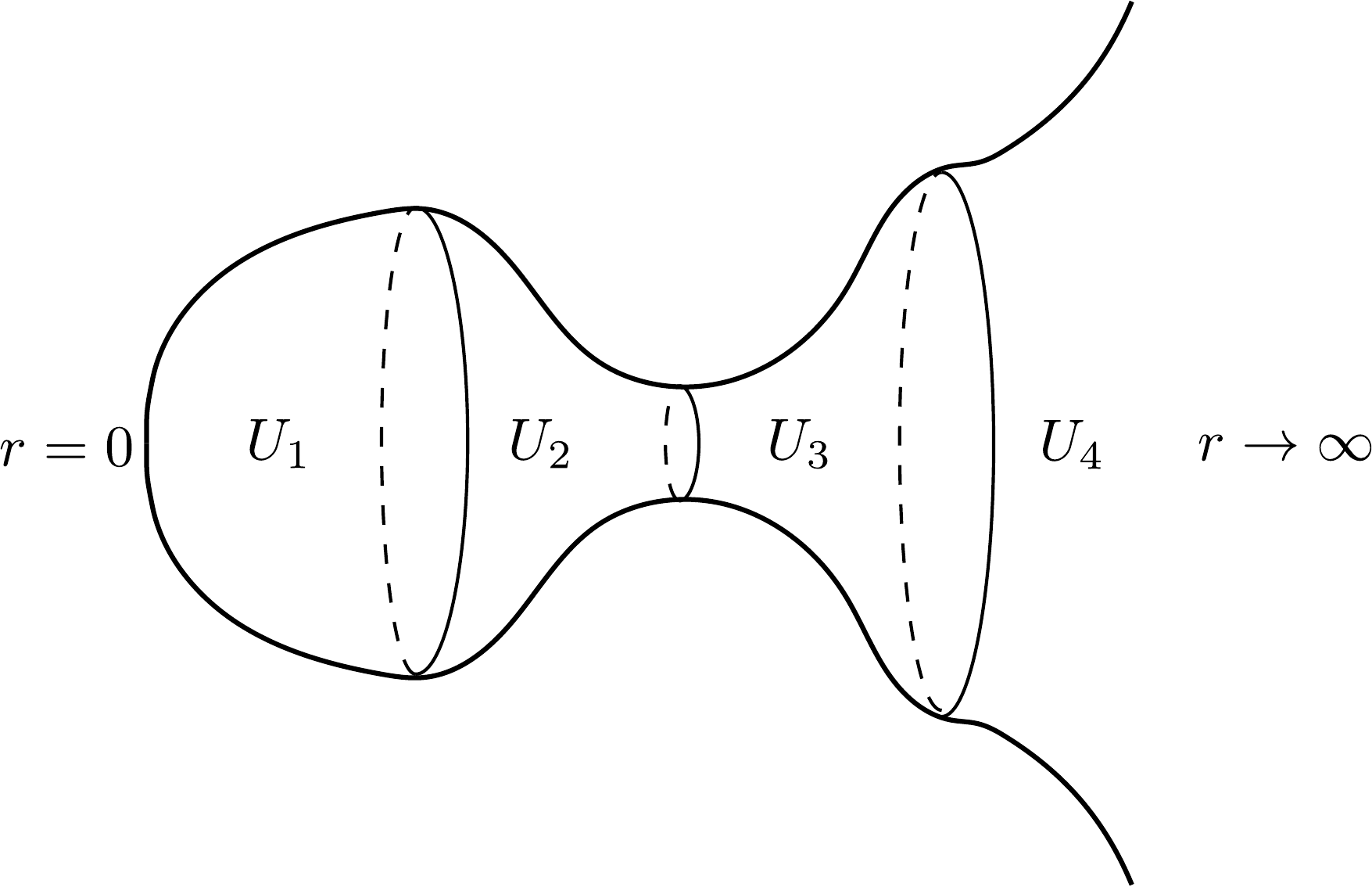}
    \caption{An example of a $d=2$ spacelike slice $\Sigma$ covered by four coordinate patches of
    the type \eqref{eq:cancoord}, separated by every type of stationary surface.
    At these surfaces, $B=\infty$.}
\label{fig:stationary}
\end{figure}

Consider now a spherically symmetric spacelike initial data slice $\Sigma$.
Because of spherical symmetry, we can cover $\Sigma$ with coordinate patches $U_i$, such that on each patch,
we have local coordinates
\begin{equation}\label{eq:cancoord}
\begin{aligned}
    h_{\mu\nu}\dd y^{\mu}\dd y^{\nu} = B(r) \dd r^2 + r^2 \dd \Omega^2,
\end{aligned}
\end{equation}
where $\dd\Omega^2$ is the round metric $S^{d-1}$.
These coordinates break down at some given radius $\hat{r}$ if and only if
$B(\hat{r})=\infty$. $B(\hat{r})=0$ can be shown to correspond to curvature
singularity for $h_{ab}$, which we do not consider. By computing
the mean curvature of a sphere of radius $r$ within $\Sigma$, we can see that $B=\infty$
precisely corresponds to the sphere having vanishing mean curvature \cite{EngFol21a}, meaning
that the sphere is locally minimal, maximal, or saddle-like within $\Sigma$. See
Fig.~\ref{fig:stationary}.

Next, assume furthermore the mean extrinsic curvature in spacetime vanishes: $K\indices{_a^a}=0$. For nonpositive cosmological constant, this usually means
that $\Sigma$ is a maximal volume slice, while for a positive cosmological
constant, it can also mean that $\Sigma$ has minimal volume. The extrinsic curvature is conveniently parametrized by a single
function $\mathcal{K}(r)$ as\footnote{There is only one function's worth of degrees of
freedom in $K_{ab}$, thanks to spherical symmetry and $K\indices{^a_a}=0$.}
\begin{equation}
\begin{aligned}
   K_{\mu\nu}\dd y^{\mu}\dd y^{\mu} =  \mathcal{K}(r) \left[B(r) \dd r^2 -
    \frac{ r^2 }{ B(r)(d-1)  } \dd \Omega^2\right].
\end{aligned}
\end{equation}
$\mathcal{K}(r)$ is simply the $rr$-component in an orthonormal basis, and so
must be everywhere finite, otherwise $\Sigma$ would have a singular embedding in
spacetime. Furthermore, on a slice $\Sigma$ whose embedding in spacetime does
not have any kinks, $\mathcal{K}(r)$ must be continuous across coordinate
patches.

It turns out that our analysis will greatly simplify if we introduce the function $\omega(r)$ as 
\begin{equation}\label{eq:omegadef}
\begin{aligned}
    B(r) = \frac{ 1 }{ 1 + \frac{ r^2 }{ L^2 } - \frac{ \omega(r) }{ r^{d-2} } }.
\end{aligned}
\end{equation}
In terms of this variable, a stationary surface is characterized by 
\begin{equation}
\begin{aligned}
    0 =  1 + \frac{ \hat{r}^2 }{ L^2 } - \frac{ \omega(\hat{r}) }{
        \hat{r}^{d-2} }.
\end{aligned}
\end{equation}
When we patch together two coordinate systems at some $r=\hat{r}$,
$\omega$ must match in the two patches, since $\omega(\hat{r})$ is uniquely
fixed by $\hat{r}$. Next, we have that in a patch containing conformal infinity, where we can take $r\rightarrow \infty$,
that \cite{EngFol21a}\footnote{This relies on
$K\indices{^a_a}=0$, or that $K\indices{_a^a}\rightarrow 0$ sufficiently fast at
infinity.} 
\begin{equation}
\begin{aligned}
    M = \frac{ (d-1)\text{Vol}[S^{d-1}] }{ 16\pi G_N }\omega(\infty),
\end{aligned}
\end{equation}
where $M$ is the ADM mass, or the AdS analogue. We refer to $\omega$ as the
Hawking mass.\footnote{There are two versions of the Hawking mass: the
Riemannian (also known as the Geroch-Hawking mass) \cite{Ger73,JanWal77,ChrSim01,Wan01} and the Lorentzian
versions \cite{MisSha64,Haw68,Hay98,BraHay06}, with the former not being
directly sensitive to $K_{ab}$. $\omega$ is the Riemannian version, suitably
generalized to $d \neq 3$ and $\Lambda \neq 0$, in the spacial case of spherical symmetry.
}

Computing the constraints in our setup, we find the equations 
\begin{align}
    (d-1) \frac{ \omega'(r) }{r^{d-1}} &= \left(1 +\frac{ r^2 }{ L^2 }- \frac{
        \omega }{ r^{d-2} }\right)\phi'(r)^2 + \dot{\phi}(r)^2 +\frac{ d }{ d-1
    }\mathcal{K}(r)^2, \label{eq:Escal}\\
    \frac{ \dd }{ \dd r }\left[r^{d}\mathcal{K}(r)\right] &=
    r^{d}\phi'(r)\dot{\phi}(r),\label{eq:Jscal}
\end{align}
where the first equation is the Hamiltonian constraint, and the second the
momentum constraint. Since each
coordinate patch has its own set of functions $\{\omega, \mathcal{K}, \phi,
\dot{\phi}\}$, we will sometimes write $\omega_{U_i}$ to indicate the $\omega$-function
on the $i$-th patch, and similarily for other quantities. To obtain the solution
on all of $\Sigma$, we solve these equations on each patch, matching $\omega,
\mathcal{K}$ across each junction. In a spacetime with two
asymptotic regions, there are two integration constants, corresponding to $\omega$ and $\mathcal{K}$
at a single value of $r$. These are the purely gravitational degrees of freedom
in spherical symmetry. Without matter, where every solution is (AdS-)Schwarzschild
by the Birkhoff-Jebsen theorem \cite{Jeb21,Bir23,VojRav05,SchWit10}, these
numbers determine $M$ and $t_L + t_R$, where $t_L, t_R$ are the times on the left and right conformal
boundaries at which $\Sigma$ is anchored. With only one conformal boundary, or
in a spacetime without any conformal boundary, $r=0$ is present on at least one
coordinate patch. The spacelike nature of $\Sigma$ ($B>0$) then requires that
$\omega(r=0)=0$, while smoothness of the embedding of $\Sigma$ in spacetime
requires $\mathcal{K}(r=0)=0$. 

Now come the important observations. Since we are on a spacelike slice, we must have $B(r)>0$, meaning that
the prefactor of $\phi'(r)^2$ term in \eqref{eq:Escal} is positive, and thus
also the full right hand side. We
conclude that
\begin{equation}
\begin{aligned}
    \omega'(r) \geq 0.
\end{aligned}
\end{equation}
$\omega$ is monotonically non-decreasing as we move towards
increasing $r$. This conclusion does not rely on our particular theory. It is true as
long as we have the WEC \eqref{eq:WEC}.
It is the monotonicity of $\omega$ that makes excitations in gravity more difficult
to screen.  However, even with one conformal boundary, the direction of increasing $r$ does not
always point towards the conformal boundary. This is at the heart of all our
subsequent results. It means that adding matter in a patch of spacetime were $r$ is
decreasing towards the boundary can decrease the ADM mass, despite the fact that
the matter has positive local energy density.

We are now going to use Eqs.~\eqref{eq:Escal} and \eqref{eq:Jscal} to study subregion independence and
dressing of gravitational perturbations. This requires us to study the linearized versions of
\eqref{eq:Escal} and \eqref{eq:Jscal} around some background geometry. We will
restrict to spherical perturbations preserving $K^a_a =0$ in our explicit analysis, and give 
arguments that our conclusions are unchanged under general perturbations.
Now, the scalar matter corresponds to completely free data; thus we can organize our
perturbative expansion as follows:
\begin{equation}
\begin{aligned}
    \phi &= \phi_{0}(r) + \kappa \delta \phi(r), \\
    \dot{\phi} &= \dot{\phi}_{0}(r) + \kappa \delta \dot{\phi}(r), \\
    \omega &= \omega_0(r) + \sum_{i=1}^{\infty} \kappa^{i}\delta_{i}\omega(r),
    \\
    \mathcal{K}&= \mathcal{K}_0(r) + \sum_{i=1}^{\infty}
    \kappa^{i}\delta_{i}\mathcal{K}(r),
    \\
\end{aligned}
\end{equation}
where $\delta \phi, \delta \dot{\phi}$ is free data, and with $\delta_i \omega, \delta_i
\mathcal{K}$ determined by the constraints. $\kappa$ is the small
perturbative parameter controlling our expansion.\footnote{Morally, we could think
of $\kappa$ as $\sqrt{G_N}$, although in the purely classical theory, there
is no preferred universal scale. If we were to sum the series, $\kappa$
should be adjusted according to the scales of the background solution.} We could also adjust $\delta \phi, \delta
\dot{\phi}$ more finely at higher orders, but we will not need this.

Note that the fully non-perturbative solutions of \eqref{eq:Escal}, \eqref{eq:Jscal} can be written
down exactly as nested integrals over the sources $\phi(r), \dot{\phi}(r)$ (see
for example \cite{EngFol21a}), but we will not need these solutions.

\subsection{Subregion independence in asymptotically flat and AdS spacetimes}
We have already shown how a compact region in the center of AdS or Minkowski
depends on an annulus around infinity. From the perspective of spherically
symmetric perturbations, this follows from integrating \eqref{eq:Escal} from
$r=0$ to $r=\infty$ and using $\omega(r=0)=0$, giving that
\begin{equation}
\begin{aligned}
    \omega(\infty) = \omega(\infty) - \omega(0) = \int_{0}^{\infty}\dd r
    (\text{positive}) > 0.
\end{aligned}
\end{equation}
Thus, let us now consider more interesting geometries. For
notational convenience we take $d=3$ and a vanishing cosmological constant -- 
the analysis is virtually identical for AdS and/or $d\neq 3$. We will treat the
$\Lambda > 0$ case separately.

\subsubsection*{Schwarzschild}
\begin{figure}
\centering
\includegraphics[width=0.8\textwidth]{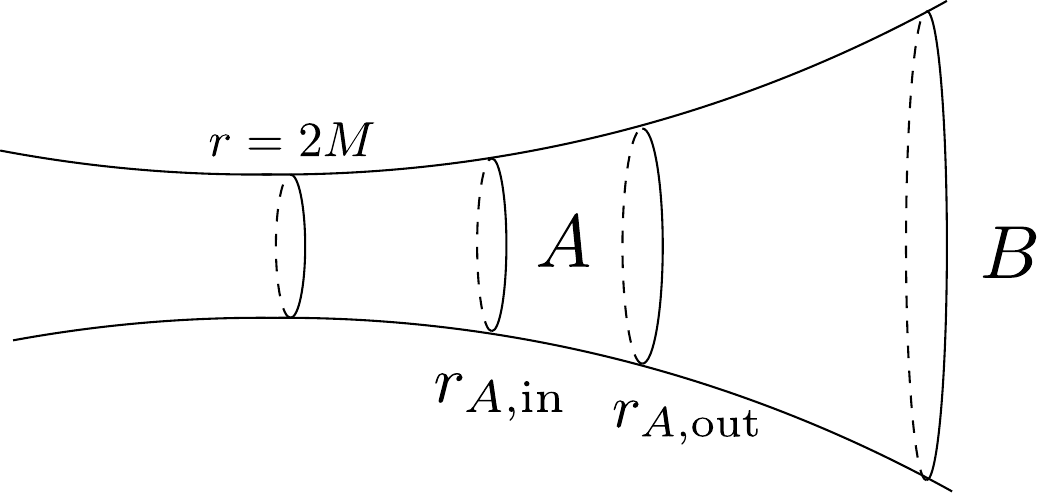}
    \caption{A bounded annular region $A$ on a $t=0$ slice in Schwarzschild, with one
    spatial dimension suppressed. $A$ and $B$ are dependent regions.}
\label{fig:schw}
\end{figure}
Let us consider the simplest non-trivial background: Schwarzschild with mass $M$. 
Let $\Sigma$ be a canonical $t=0$ slice, which has $K_{ab}=0$. Let $B$ be an infinite annulus
containing right infinity, and let $A$ be a finite annulus in the right region, restricted
to $r\in [r_{A, \text{in}}, r_{A, \text{out}}]$, with $r_{A, \rm{in}}>2M$. See
Fig.~\ref{fig:schw}. The perturbed leading order constraints read\footnote{Since
the leading order perturbation of $\delta T_{ab}$ is $\mathcal{O}(\kappa^2)$, we
do not consider $\delta_1 \omega\neq 0$.}
\begin{equation}\label{eq:omega2}
\begin{aligned}
    2 \delta_2 \omega'(r) &= r^2 f(r)\delta
    \phi(r)^2 + r^2\delta \dot{\phi}^2,  \\
    \frac{ \dd }{ \dd r  }\left(r^{3}\delta_2 \mathcal{K}(r)\right) &=
    r^{3}\delta \phi'(r)\delta \dot{\phi}(r),
\end{aligned}
\end{equation}
with $f(r)=1-2M/r$. The free data in $A$ are the functions $\delta \phi(r),
\delta \dot{\phi}(r)$ for $r \in [r_{A, \text{in}}, r_{A, \text{out}}]$, and two numbers, corresponding
to $\delta_2 \omega, \delta_2 \mathcal{K}$ 
a single value of the radius in $A$. These two numbers are the only spherically
symmetric gravitational degrees of freedom. To test for independence, we are free to consider a perturbation with $\delta_2
\omega(r_{A, \text{in}})=\delta_2
\mathcal{K}(r_{A, \text{in}}) =0$, and add some matter in $A$. With the choice 
$\delta_2\omega(r_{A, \text{in}})=\delta_2 \mathcal{K}(r_{A, \text{in}})=0$,
we are choosing $\delta S|_A$ such that the new gravitational field
sourced by $\delta \phi$ is forced to travel to the right. 
Then, integrating \eqref{eq:omega2} from $r_{A,\text{in}}$ to $r=\infty$ gives that $\delta_2 \omega(\infty)>0$
on the right, and so $A$ and $B$ are dependent. The background has no matter
that we can remove outside $A$ to screen the perturbation. One could wonder if a non-symmetric perturbation outside $A$ could be
used to screen the perturbation, but in this case, we can appeal to
a rigorous theorem, the so-called Riemannian Penrose inequality
\cite{HuiIlm01,Bra01,BraLee09},\footnote{In the AdS case, this inequality is
still conjectural, and only partial proofs exist
\cite{LevFre12,GeGuo13,HusSin17,EngFol21a,Fol22}.
} to
show that no perturbation whatsoever can save us.\footnote{We do assume that $K^a_a=0$ is preserved under
these perturbations, but is not a severe restriction, since small perturbations
should not break the existence of a slice with maximal volume. So even if we
perturbed away from $K^a_a=0$, there should be a nearby slice in spacetime where
$K^a_a=0$, so we can always take our perturbation to be a combination of the
perturbation that changes spacetime, together with a diffeomorphism that
moves us to this slice.} This theorem says that if we have an asymptotically
flat initial dataset with $K\indices{_a^a}=0$, with one asymptotic end with mass
$M$, and with an inner boundary
$\sigma$ that is an outermost minimal surface, then
\begin{equation}
\begin{aligned}
    M \geq \sqrt{\frac{ \text{Area}[\sigma] }{ 16 }},
\end{aligned}
\end{equation}
with equality if and only if we are identically Schwarzschild. 
Our chosen perturbation in $A$ was compatible with keeping the geometry at
the minimal surface fixed, and so if a non-symmetric dressing that did not
reach $B$ existed, it would violate the rigidity statement of the Riemannian
Penrose inequality. So graviton degrees of freedom cannot make $A$ and $B$
independent.

\begin{figure}
\centering
\includegraphics[width=0.8\textwidth]{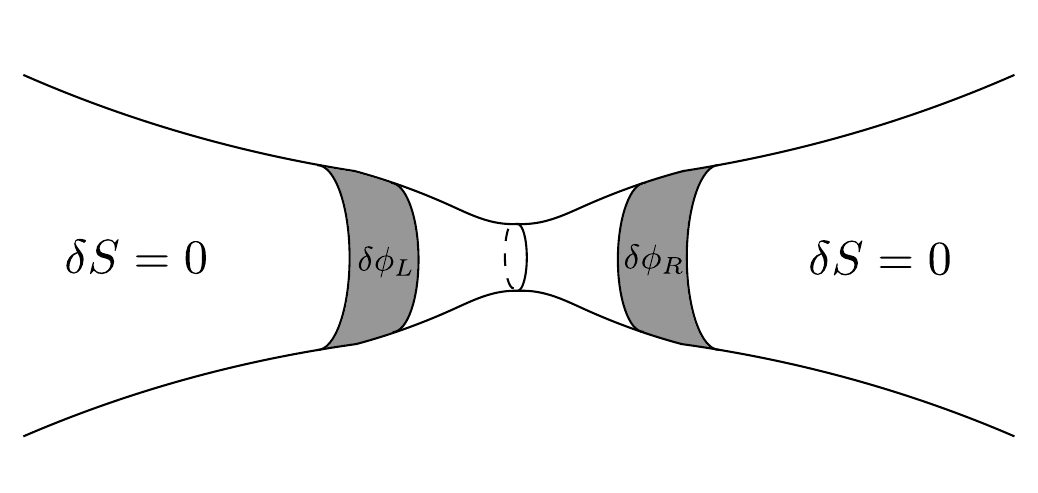}
    \caption{A perturbation $\delta \phi_R$ on the right side dressed to
    a perturbation on the left side, leaving the initial data both asymptotic
    regions unchanged. This causes the area of the minimal surface to
    decrease.}
\label{fig:schw2}
\end{figure}

We thus see that any bounded
region $A$ lying strictly in the right exterior is dependent on right infinity.
Is $A$ dependent on left infinity? To test this, again pick a perturbation where we add some matter in
$A$. However, we now use our gravitational degree of freedom to set
$\delta_2 \omega(r_{A, \text{out}})=\delta_2 \mathcal{K}(r_{A, \text{out}})=0$ instead,
forcing the new gravitational field lines sourced by $\delta \phi$ to travel to the left. This seems dangerous: will we now be forced to dress to left infinity?
The answer is no! We can screen the perturbation from left infinity by
\textbf{adding} more matter, provided we add it to the left of the bifurcation
surface. Integrating the constraints, and matching $\delta_2 \omega, \delta_2
\mathcal{K}$ across the coordinate systems, we find that the functions on the
left side are\footnote{The reader
might be worried that $\delta_2 \omega \neq 0$ at $r=2M$, since the $rr$-metric
perturbation there reads, in our gauge, $\delta_2 h_{rr} = r\delta \omega
(r-2M)^{-2}$, which
diverges at $r=2M$. However,
this is simply a gauge artifact that can be fixed with the addition of a term
$\nabla_{(a}\xi_{b)}$. It is simply reflecting the fact the location of the
minimal surface, and thus coordinate breakdown location of our gauge, has moved by a perturbative amount. }
\begin{align}
    2\delta_2 \omega_{L}(r) &= 
    \int_{2M}^{r}\dd \rho
    \rho^2 \left[f(\rho)(\delta \phi'_{L})^2
    + (\delta \dot{\phi}_{L})^2 \right] 
    +
    \int_{r_{A, \text{out}}}^{2M}\dd \rho
    \rho^2 \left[f(\rho)(\delta \phi'_{R})^2
    + (\delta \dot{\phi}_{R})^2 \right], \label{eq:omega2sch} \\
    \delta_2 \mathcal{K}_{L}(r) &= \int_{2M}^{r} \dd \rho \rho^3 \delta
    \phi'_{L}(r)\delta \dot{\phi}_{L} + \int_{r_{A,\text{out}}}^{2M} \dd \rho
    \rho^3 \delta \phi'_{R}(r)\delta \dot{\phi}_{R}.
\end{align}
The crucial thing to note in these equations is that the last term in
\eqref{eq:omega2sch} is negative, caused by that fact that $r$ increases in
opposite directions on each side of the bifurcation surface. Locally positive energy
densities on one-side of an extremal surface
contribute negatively to the Hawking (and thus ADM) mass on the other
side.
Thus, we see that we can always pick $\delta \phi_L, \delta
\dot{\phi}_L$ such that $\delta_2 \omega_L(r)=0, \delta_2 \mathcal{K}_L(r)=0$
for all $r>2M+\epsilon$ for any $\epsilon>0$ (assuming $\epsilon$ does not scale
with $\kappa$ to a positive power). So a positive energy density shell can be
dressed to another positive energy density shell, provided they are separated by
an extremal surface. See Fig.~\ref{fig:schw2}. As advertised earlier, now gravity behaves more like
electromagnetism. Positive energy densities seen from the other side of an
extremal surface behaves as a negative energy density.

What about perturbations in $A$ that are not spherically symmetric, but where we
keep the solution in $A$ fixed near the rightmost boundary, so that we are
forcing the gravitational field to change towards the left.
Can these perturbations be screened from left infinity? It is physically
quite clear that they can, although we will give an argument rather a proof. First, using a technique known as inverse mean curvature
flow, which we explain in Sec.~\ref{sec:IMC}, it can be shown for $d=3$ that non-spherical
additions of matter on the right will
again contribute negatively to the mass on the left. So again we can simply add matter
on the left to bring this back up, keeping left the mass unchanged. Of course,
now we might also have to cancel the momentum and angular momentum,
but this is much easier since these are quantities without a preferred sign. Let
us assume we added some amount of angular momentum in $A$ that is backreacting
leftwards. We might worry that
the required matter or gravitons we need to add to cancel this angular momentum
forces us to overshoot the mass, causing a left mass increase. But this can
always be avoided. Rather than cancelling the angular momentum by adding matter on
the left, we can add it on the right side, between $A$ and the minimal surface,
so that we bring the angular momentum closer to zero, while at the same time decreasing the
mass. Thus, as long as $A$ is slightly separated from the bifurcation surface, so that we have the flexibility of perturbing on both sides of
it, $A$ ought to be independent of $B$.

What we see from this discussion is that gravity appears to have a classical
version of the split property across the Schwarzschild bifurcation surface. 
Let $A$ and $B$ be the regions
$r\geq2M+\epsilon$ on the left and right, respectively, with $\epsilon>0$ not
scaling with $\kappa$. By the discussion above, we see that $A, B$ are likely
independent of each other. However, it is
crucial that $\epsilon$ is non-zero, 
and in fact non-perturbatively large, so that we can always fit enough matter
without leaving perturbation theory. Thanks 
to the fact that the small band between $A, B$ contains an extremal surface, we
have access to perturbations with all combinations of signs of charges within this band. 
We will see later that this property is a feature of general extremal surfaces.

\subsubsection*{A spacetime with a lump of matter}
\begin{figure}
\centering
\includegraphics[width=0.5\textwidth]{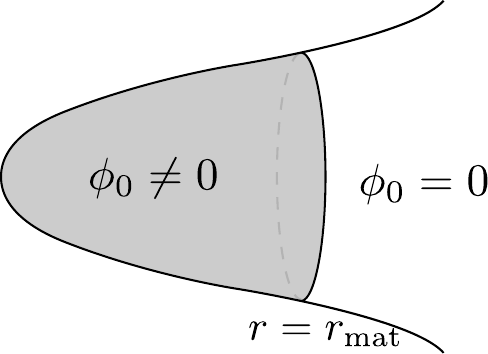}
    \caption{A one-sided spacetime with a lump of matter around $r=0$. We argue
    that a ball $A$ with radius $r_A$
    is independent from an annulus $B$ around infinity if and only if $r_A<r_{\rm
    mat}$ }
\label{fig:onesided}
\end{figure}

Consider now a one-sided spherically symmetric geometry at a moment
of time-symmetry, $\mathcal{K}_0=\dot{\phi}_0=0$. Assume that it has no
stationary surfaces, so that one coordinate patch covers everything. Next, we
take there to be a lump of matter $\phi_0(r)$ with compact
support on $r\in[0, r_{\rm mat}]$. See Fig.~\ref{fig:onesided}. We demand that
$\phi_0'(r)$ is non-zero at least
in $r\in [r_{\rm
mat}-\epsilon, r_{\rm mat})$ for some $\epsilon>0$. 

Let $A$ be a ball with radius $r_A$, and take $B$ to be an annulus
around infinity. Consider turning on any scalar
field profiles $\delta \phi(r), \delta \dot{\phi}(r)$ in $A$. The perturbative
constraints are now non-trivial at first order in $\kappa$, and read
\begin{equation}\label{eq:matterball}
\begin{aligned}
    \delta_1 \omega'(r) &= r^2\left[1- \frac{ \omega_0(r)
    }{ r }\right]\phi_0'(r)\delta \phi'(r) - \frac{ 1 }{ 2
    }r\delta_1\omega(r)\phi_0'(r)^2, \\
    \frac{ \dd }{ \dd r }\left(r^{3}\delta_1 \mathcal{K}\right) &= r^3
    \phi'_0(r)\delta \dot{\phi}(r).
\end{aligned}
\end{equation}
Can we then extend the perturbation to $r>r_A$ such that no backreaction leaks to infinity? If $r_A<r_{\rm mat}$, then the answer is yes. The
first order constraint solutions are
\begin{align}
    \delta_1 \omega(r) &= e^{-\frac{ 1 }{ 2
    }\int_{0}^{r}\dd z z \phi_0'(z)^2}\int_{0}^{r} \dd \rho\rho^{2}e^{\frac{ 1 }{ 2
    }\int_{0}^{\rho}\dd z z \phi_0'(z)^2}\left[1-
    \frac{ \omega_0(\rho)
    }{ \rho }\right]\phi_0'(\rho)\delta \phi'(\rho) \label{eq:omega1sol}\\
    \delta_{1}\mathcal{K}(r) &= \frac{ 1 }{ r^3 }\int_{0}^{r}\dd \rho \rho^3
    \phi_0'(\rho) \delta \dot{\phi}(\rho) \label{eq:K1sol}
\end{align}
We now see that neither integral has fixed sign, 
so as long as $r_A<r_{\rm mat}$\footnote{And $r_A - r_{\rm
mat}=\mathcal{O}(\kappa^0)$.} we can always pick $\delta \phi, \delta\dot{\phi}$ 
on $(r_A, r_{\rm mat}]$ so that there is some $\hat{r} \in (r_A, r_{\rm mat}]$
such that
\begin{equation}
\begin{aligned}
    \delta_1\omega(r\geq \hat{r})=0,\quad \delta_1 \mathcal{K}(r\geq \hat{r})
    = 0.
\end{aligned}
\end{equation}
Physically, what has
happened is that we have removed some matter from the background shell in order
to avoid having the backreaction of our newly added matter leak out to infinity.
Since we are considering infinitesimal perturbations, the background reservoir is
effectively infinitely large, so we can keep doing this to any order in perturbation theory.
All spherically symmetric perturbations in $A$ can be dressed to the background lump of
matter. Similarly, if we alter the mass in $B$, we can always add or remove some
of the matter in the shell lying in $C$ to compensate, keeping $A$ fixed. 
Assuming we can dress non-symmetric perturbations to the lump as well,
we thus find that $A, B$ now are independent. It is physically reasonable that it should
also be possible to dress these perturbations to $B$. If our non-spherical perturbation
in, say, $A$, adds some momentum or angular momentum, we can just add more moving or
rotating matter between $A$ and $B$ to cancel these charges, since they do not have a
preferred sign. The only real worry is that this matter will
increase the energy at the same time, but since we are working perturbatively, we have an
infinite background reservoir of matter, so we can always remove a bit of matter
from the lump to keep the energy fixed. So the background lump of matter restores
independence of the ball $A$ from infinity, and perturbations in $A$ need not be
dressed to infinity.

If we instead took $r_A>r_{\rm mat}$, independence breaks down. This is
because the spacetime looks like Schwarzschild for $r>r_{\rm mat}$, and 
if we pick a perturbation on $A$ that changes the region $(r_{\rm mat}, r_A]$,
but keeps everything fixed for $r \in [0, r_{\rm mat}]$, then we
have specified a perturbation where no matter from the background is allowed to be removed, and we have
no choice but to let the backreaction leak to infinity. This time there is no
minimal surface to dress to. Non-symmetric perturbations in $C$ cannot save us, since 
these would constitute all-order compactly supported perturbations of pure Schwarzschild in one exterior, and these perturbations could
be used contradict the rigidity-part of the Riemannian Penrose inequality.

This example illustrated dressing of perturbations to background matter. It was
important that this background matter was not just perturbatively small, since
then the argument would break down -- there would exist perturbations that could
not be dressed to matter, since the added matter then exceed what we might have
the ability to remove. At the level of the full perturbative series, the radius
of convergence of $\kappa$ thus has to be less than infinity. 

\subsubsection*{Schwarzschild in pure gravity}
\begin{figure}
\centering
\includegraphics[width=1.0\textwidth]{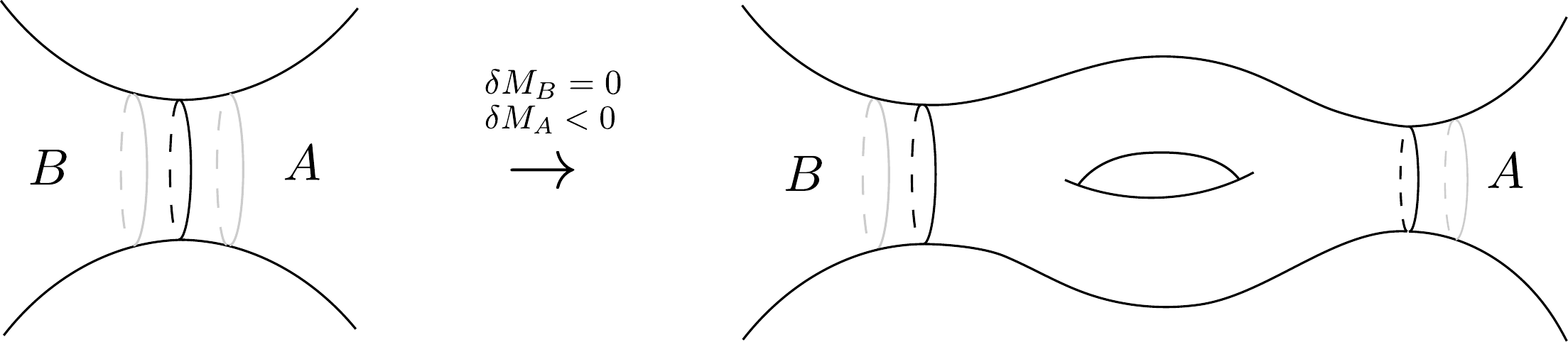}
    \caption{Using pure gravity degrees of freedom to change the mass of
    the BTZ region on the $A$-side without changing it on the $B$-side. The right geometry looks identical to BTZ in $A$ and $B$. }
\label{fig:BTZ}
\end{figure}

We already discussed subsystem independence in Schwarzschild, but we crucially relied on matter. What happens in vacuum
gravity? Consider again a canonical $t=0$ slice of (AdS-)Schwarzschild, and let 
$A$ and $B$ be the regions $r\geq 2M + \epsilon$ on the
right and left, respectively.  Consider now picking a perturbation 
$\delta S|_A$ that corresponds to simply deforming $A$ to Schwarzschild with a lower mass. 
 Pure gravity has no local spherically symmetric degrees of freedom, so
we can clearly not hide this perturbation from $B$ within spherical symmetry.
However, it appears likely that the perturbation can be hidden
from $B$ if we break spherical symmetry, so that we unlock the graviton degrees of
freedom. Let us see how this happens in a similar situation: the BTZ black
hole. Unlike higher dimensions, we do not have local degrees of freedom. However, vacuum
gravity still possesses topological degrees of freedom. Once we leverage these, it
is well known (see for example \cite{AmiBen97,Bri98}) that we can reduce the BTZ mass in $A$
without changing $B$ by filling in a higher-genus surface between $A$ and $B$ --
see Fig.~\ref{fig:BTZ}. Of course, in this case, the change
is not perturbative, since we changed the topology, so pure 3d gravity at best
has a non-perturbative notion of independence. Nevertheless, the example is
brought up to make it
less surprising that pure gravity degrees of freedom might do the job for $d\geq
3$. In fact, there are mathematical results in GR supporting this
possibility. As long as we are allowed to change the geometry at the bifurcation surface, we are not
constrained by the Riemannian Penrose inequality, and Corvino and Schoen
\cite{Cor00,CorSch06} have shown that being identically Schwarzschild in a
neighbourhood of infinity is not a rigid feature.\footnote{Specifically, they proved
that for any asymptotically flat vacuum solution $(h_{ab}, K_{ab})$ of the
constraints on $\mathbb{R}^3$, and for any choice of compact subset
$\Omega\subset\mathbb{R}^3$, there exists new vacuum initial data $(h'_{ab},
K'_{ab})$ that looks identical to Kerr near infinity, and which agrees
with $(h_{ab}, K_{ab})$ on $\Omega$.} Furthermore, by Theorem 1.1 in
\cite{Hin23}, Schwarzschild is known to support first order
pure gravity perturbations of compact support. Thus, it is a possibility that any perturbation on
$A$ can be shielded from $B$ by utilizing pure gravity degrees of freedom in $C$. 

\subsection{Subregion independence in de Sitter}
\begin{figure}
\centering
\includegraphics[width=1.0\textwidth]{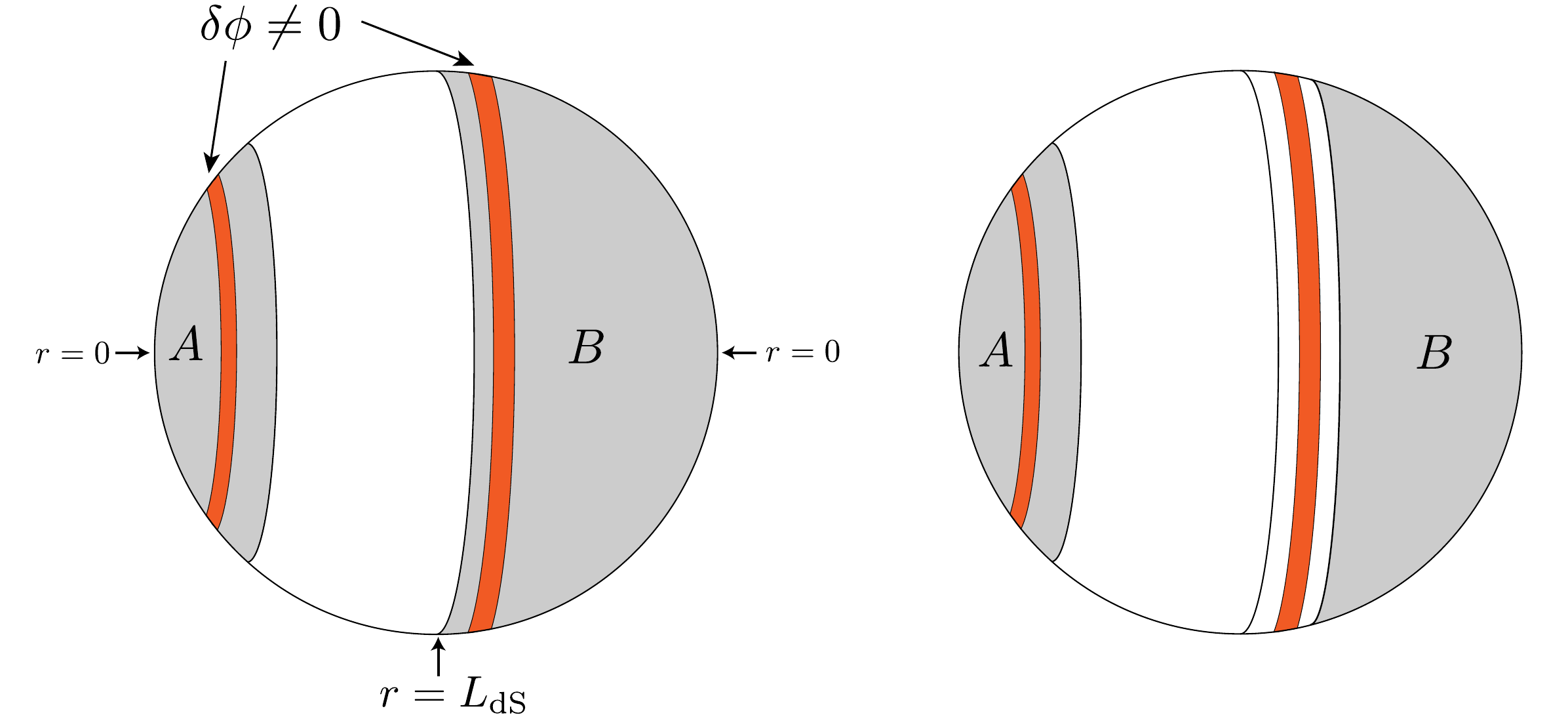}
    \caption{A perturbation of a minimal slice of de Sitter, which is just a
    round sphere of radius $L_{\rm dS}$. The two shells are
    dressed to each other, so the solution looks like pure de Sitter near the
    poles of $A$ and $B$. At the level of spherical symmetry, adding a shell to
    the one static patch requires the addition of one to the other. 
    Thus it is possible that $A$ and $B$ are dependent in the left scenario. On
    the right, we can screen perturbations in $A$ from $B$ by adding matter
    around the maximal surface.
    }
\label{fig:dS}
\end{figure}

Let us now treat a case with positive cosmological constant. Let
the background be a minimal slice of dS$_{4}$, which has $K_{ab}=0$, and which can be
covered by two coordinate patches $U_L, U_R$, each having a range $r\in[0,
L_{\rm dS}]$. The metric on each patch is
\begin{equation}
\begin{aligned}
    \dd s^2 = \frac{ \dd r^2 }{ 1 - \frac{ r^2 }{ L_{\rm dS}^2 }} + r^2
    \dd\Omega^2,
\end{aligned}
\end{equation}
i.e. $\omega_0(r)=0$. Together these two patches make up a round sphere of
radius $L_{\rm dS}$.
The spherically symmetric constraints are non-trivial at $\mathcal{O}(\kappa^2)$, and given by \eqref{eq:omega2}, except now $f(r)=1-r^2/L_{\rm dS}^2$.
At $r=0$, we as usual
need that $\delta_2 \omega(0)=0$. Integrating \eqref{eq:omega2} and using that
$\delta_2\omega$ is continuous at $r=L_{\rm dS}$, all perturbations are constrained to satisfy
\begin{equation}\label{eq:deSitterChange}
\begin{aligned}
\int_{0}^{L_{\rm dS}}\dd \rho
    \rho^2 \left[f(\rho)(\delta \phi'_{L})^2
    + (\delta \dot{\phi}_{L})^2 \right] = \int_{0}^{L_{\rm dS}}\dd \rho
    \rho^2 \left[f(\rho)(\delta \phi'_{R})^2
    + (\delta \dot{\phi}_{R})^2 \right] 
\end{aligned}
\end{equation}
This is just a special case of the first law of thermodynamics for positive
cosmological constant \cite{GibHaw77} applied to spherical perturbations of pure dS.
Since $U_L$ and $U_R$ are Cauchy slices for the left and right static patches, we
see that within the domain of spherical
symmetry, it is impossible to perturb one
static patch without also perturbing the other.
See Fig.~\ref{fig:dS}. It is easy to show the same result for all $d\geq 2$. Thus, it seems plausible in dS$_{d+1}$ for $d\geq 2$, any subregion of a given 
static patch depends on the full opposite static patch. However, we have no proof 
that we cannot break spherical symmetry on the right and leverage pure vacuum
gravity degrees of freedom to shield this perturbation from the left. As we discuss in
Sec.~\ref{sec:dS}, there very likely exist at least some non-symmetric
non-trivial initial data that is strictly localized to one static patch.
Either way, $A$ is probably independent of any $B$ that is a strict
subset of the opposite static patch, provided $B$ is such that $C$ contains a
neighbourhood of the cosmological horizon. In this case, we  
can satisfy \eqref{eq:deSitterChange} by adding matter in a small
neighbourhood around the cosmological horizon, as illustrated in
Fig.~\ref{fig:dS}. 

In recent work, a von Neumann algebra of diffeomorphism invariant
operators was constructed for a static patch of de Sitter that contains an observer
\cite{ChaLon22}. This algebra consisted of operators dressed to the observer.
It was found that in order to have a consistent description 
also describing the opposite static patch, it was necessary to include an observer in the
second patch as well. While our setup is different, it is worth pointing out
that an observer ought to backreact on the geometry, and the Hamiltonian constraint
forbids backreaction in a single static patch (in spherical symmetry). If the
observer were to travel on a worldline, this perturbation preserves spherical symmetry,
and so some backreaction must thus be added in the other patch. Adding a second
observer solves this. 
In our setup, it would however be appropriate to treat an observer as a feature of the
background, rather than a perturbation, so that dressed observables act as perturbations
with respect to the observer. To treat this case, and to get a better sense of
when we can deform a single static patch, we prove a stronger theorem
in Sec.~\ref{sec:dS}, showing that for full spherically symmetric non-linear backreaction and
$d\geq 2$, it is not possible to change a single static patch. We do this by
showing that the area of the cosmological horizon must be reduced by any of
these deformations. However, before we do this, we first study gravitational
independence at the more general level.

\section{Dressing across extremal and trapped
surfaces}\label{sec:general}
Above we studied examples of spherically symmetric backgrounds, and we argued
that at moments of time-symmetry, we can always dress
perturbations across minimal or maximal surfaces, at least when the theory has
matter. The assumption of time-symmetry and spherical symmetry is however 
restrictive. Furthermore, it would be more useful to characterize the surfaces
that we are able to dress across in terms of their spacetime properties, rather
than their properties on a particular slice. We are now going to argue that we
can dress across extremal surfaces and generic trapped surfaces.  We start with
general spherically symmetric spacetimes and perturbations in
Secs.~\ref{sec:spheresym},\ref{sec:lorentzian}. Then in Sec.~\ref{sec:IMC}, we remove symmetry
assumptions in the case of four spacetime dimensions.

\subsection{Spherical symmetry}\label{sec:spheresym}
In spherical symmetry, the general constraint equations on a $K\indices{_a^a}=0$ slice
$\Sigma$ read
\begin{align}
    (d-1) \frac{ \omega'(r) }{ r^{d-1} } &= 2\mathcal{E} + \frac{ d }{ d-1
    }\mathcal{K}(r)^2, \label{eq:fullConstraint}\\
    \frac{ \dd }{ \dd r }\left(r^{d}\mathcal{K}\right) = r^d\mathcal{J}_{r}.
\end{align}
By virtue of $\omega'(r) \geq 0$, as we cross any minimal or maximal surface $\sigma$, we always have a flip in the
direction in which $\omega$ grows as it encounters matter. So if we add some matter on one side
of $\sigma$, we can add some matter on the other side to
compensate, keeping $\omega$ unchanged outside a neighbourhood of
$\sigma$. The assumption of time-symmetry in our examples was
irrelevant to this central point.
However, at first sight, minimal and maximal surfaces in a $K_{a}^a=0$ slice
seem like a very restricted set of surfaces. It turns out this is not true.
A simple computation shows that stationary surfaces on a $K\indices{_a^a}=0$ slice are always
either strictly (anti)trapped or extremal (see for example Eq.~(A.55) of
\cite{EngFol21a}). In a moment we are going to show a partial converse in spherical symmetry: 
maximal or minimal surfaces on $K\indices{_a^a}=0$ slices include all (spherical) extremal surfaces and
also all generic trapped surfaces, provided we do not require $\Sigma$ to satisfy $K\indices{_a^a}=0$ globally.\footnote{Allowing saddle-type surfaces, all trapped surfaces are included.} 
This will be enough for us.

\begin{figure}
\centering
\includegraphics[width=1.0\textwidth]{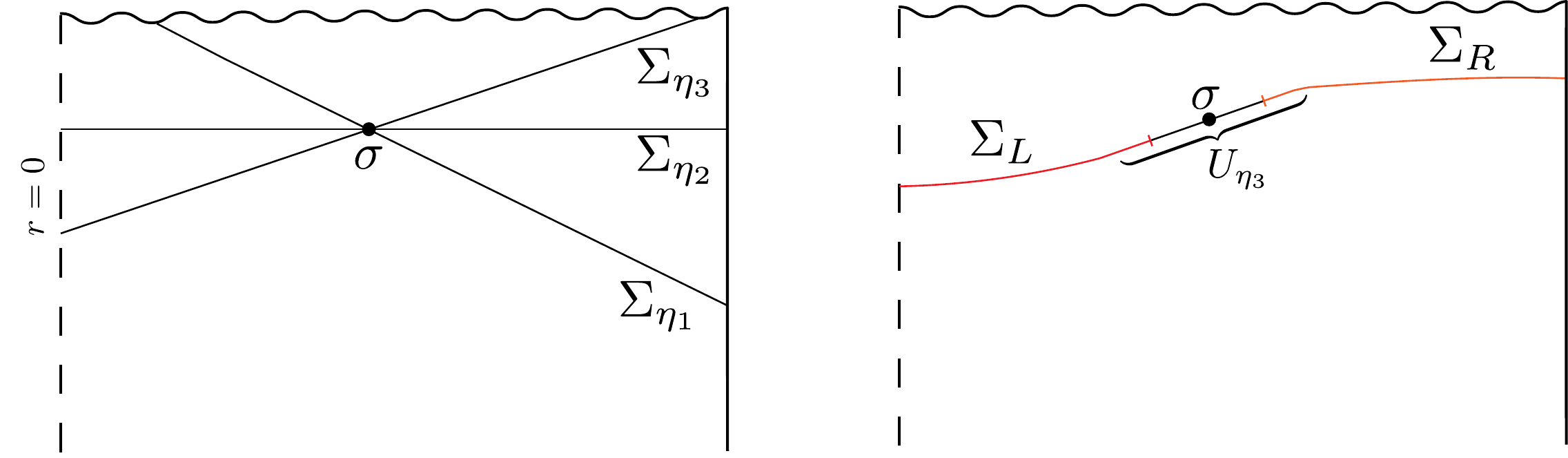}
    \caption{(Left) Three $K\indices{^a_a}=0$ slices fired off $\sigma$, with
    $\eta_3 > \eta_2 > \eta_1$.
    Only $\Sigma_{\eta_2}$ is smooth and complete. (Right) Deforming
    $\Sigma_{\eta_3}$ to a smooth complete slice, preserving a neighbourhood
    $U_{\eta_3}$ around $\Sigma$ that has $K\indices{^a_a}=0$.}
\label{fig:fired}
\end{figure}

Consider a general spherically symmetric spacetime $(N, g_{ab})$ and let
$\sigma$ be any sphere. Then there always exists a one-parameter family
$\Sigma_{\eta}$ of
spherically symmetric $K\indices{_a^a}=0$ slices locally defined in a neighbourhood around $\sigma$.
To see this, note that the equation that determines $\Sigma_{\eta}$ is a second
order ODE. This ODE is obtained by extremizing the
induced volume of $\Sigma$ with respect to a single embedding coordinate, say
$t(r)$.  One integration constant is fixed by demanding that
$\Sigma_{\eta}$ passes through $\sigma$. The second integration constant is just
the boost angle $\eta$ at which $\Sigma_{\eta}$ is ``fired off'' $\sigma$. This
is completely analogous to how we can fire a one-parameter family of radial
spacelike geodesics off some given point. Once
$\eta$ is fixed we can integrate the ODE determining the location of
$\Sigma_{\eta}$ to try to extend it to
a full slice, but generically one does not get a full smooth Cauchy slice.
See the left panel of Fig.~\ref{fig:fired}.
In a one-sided spacetime, for example, this could happen because we hit $r=0$
with the wrong boost angle, so that $\Sigma_{\eta}$ has a kink, corresponding to
$\mathcal{K}\rightarrow \infty$. Or we could fall into a singularity.
However, crucially, we will not
need a full $K\indices{_a^a}=0$ slice.  We can just terminate the integration at some
finite value and then arbitrarily continue the slice in a smooth way,  now giving up the
requirement that the mean curvature vanishes. This way we produce a
one-parameter family of Cauchy slices $\Sigma_{\eta}$, each containing $\sigma$
and each having a neighbourhood $U_{\eta} \subset \Sigma_{\eta}$ that has
$K\indices{^a_{a}}=0$. See the right panel of Fig.~\ref{fig:fired} for one such
slice.

Consider now $\sigma$ being a surface such that there exists an $\eta$ so that
 $\sigma$ is minimal or maximal on $\Sigma_{\eta}$. Let $\Sigma_{L}$ and $\Sigma_{R}$ be the parts of
$\Sigma_{\eta}$ lying to the left and right sides of $\sigma$, respectively,
except for a small neighbourhood around $\sigma$ that is included in neither --
see Fig.~\ref{fig:fired}. Thus, $\Sigma_{L}$ and $\Sigma_{R}$ are separated by an open neighbourhood
containing $\sigma$ that has $K\indices{^a_a}=0$. We now see that any spherical perturbation in
$\Sigma_{L}$ can be screened from $\Sigma_{R}$, and vice versa, by exactly the mechanism we
described earlier.\footnote{Strictly speaking, we should also consider 
perturbations that satisfy $\delta K\indices{_a^a} \neq 0$ at $\partial
\Sigma_R, \partial \Sigma_L$, so we should technically have consider the
spherically symmetric Einstein equations for general $K^a_a=0$. This leads to no
complications. This should be clear from the following Lorentzian analysis.} 

Note that $\omega(r)$ is defined everywhere on the slice by
\eqref{eq:omegadef}, but $\omega$ generally has no monotonicity properties away from $U_{\rm
\eta}$. This does not matter to the argument, however. Thus, what we now to show is that for any extremal or
generic trapped surface, there is some $\eta$ on which $\sigma$ is maximal or
minimal on $\Sigma_{\eta}$.

Let us now first assume that $\sigma$ is extremal. That means that the area of
$\sigma$ is stationary under all variations, so for every choice of $\eta$, $\sigma$ is either maximal,
minimal, or a saddle in $\Sigma_{\eta}$. Being a saddle is clearly a
fine-tuned case, and it can always be avoided. To see this, note that when the
boost angle $\eta$ approaches $+\infty$ or $-\infty$, $U_{\eta}$ gets closer and
closer to one of the two the null congruences fired off $\sigma$.  By the
focusing theorem, the area of these congruences are shrinking both to the future
and past, so once $|\eta|$ is large enough, $\sigma$ will be a locally maximal
surface on $U_{\eta}$. In the fine tuned case where the congruences
fired from $\sigma$ are stationary, we are just locally in a standard spherical black hole, in
which case we can just pick one of the other slices, where we know that $\sigma$ is minimal or
maximal.

Next, assume that $\sigma$ is a trapped or antitrapped surface, meaning that the two future
null expansions have the same sign, with both strictly nonzero. Fixing an
arbitrary zero-point of the boost angle $\eta$, let $n^a$ be
the future timelike unit normal to $\Sigma_{\eta=0}$, and let $r^{a}$ be a unit
normal to $\sigma$ that is tangent to $\Sigma_{\eta=0}$. 
Two independent future directed null normals to $\sigma$ can be taken to be
\begin{equation}\label{eq:kpm}
\begin{aligned}
    k^a_{\pm} = \frac{ 1 }{ \sqrt{2} }(n^a \pm r^a),
\end{aligned}
\end{equation}
where $k_+^a$ conventionally defines ``outwards''. 
As shown in the appendix of \cite{EngFol21a}, with this normalization of the null normals, 
the mean curvature $H_0[\sigma]$ of $\sigma$ within $\Sigma_{\eta = 0}$ reads
\begin{equation}
\begin{aligned}
    H_0[\sigma] = D_a r^a = \frac{ 1 }{ \sqrt{2} }(\theta_{+} - \theta_{-}),
\end{aligned}
\end{equation}
where the null expansions are given by $\theta_{\pm}=(g^{ab} + 2k^{(a}_+ k_{-}^{b)})\nabla_a k_b$. 
Consider now boosting our slice to $\Sigma_{\eta}$. Since $n^a, r^a$ form an orthonormal basis of
the normal bundle of $\sigma$, they transform in the canonical way under a
Lorentz boost. It is easily seen that the null normals canonically
normalized with respect to the boosted $n^a, r^a$ are
\begin{equation}
\begin{aligned}
    k_{\pm, \eta}^{a} = e^{\pm \eta}k_{\pm}^a,
\end{aligned}
\end{equation}
so the mean curvature of $\sigma$ within $\Sigma_{\eta}$ is 
\begin{equation}
\begin{aligned}
    H_{\eta}[\sigma] = \frac{ 1 }{ \sqrt{2} }(e^{\eta}\theta_{+} - 
    e^{-\eta}\theta_{-}),
\end{aligned}
\end{equation}
where $\theta_{+}, \theta_{-}$ still are the expansions at $\eta=0$.
Because $\theta_+, \theta_-$ are nonzero and have the same sign, 
we can always take $\eta = \frac{ 1 }{ 2 }\log(\theta_-/\theta_+)$, giving $H_{\eta}[\sigma]=0$. Hence, locally there is always a
$K\indices{^a_a}=0$ slice on which $\sigma$ is stationary. Assuming we do not
have the fine tuned situation where this surface is a saddle, we see that
$\sigma$ posses a slice one which the left and right sides are independent -- at
least with respect to spherical perturbations. 

\begin{figure}
\centering
\includegraphics[width=0.5\textwidth]{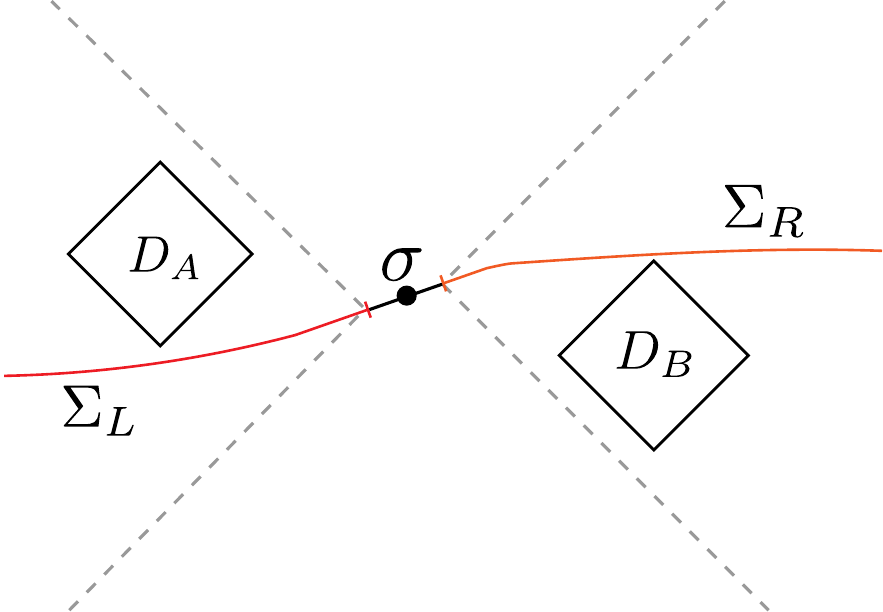}
    \caption{Two causal diamonds separated by a trapped or extremal
    surface $\sigma$.
    }
\label{fig:diamonds2}
\end{figure}

So in total, if $D_A$ and $D_B$ are two spacelike separated causal diamonds whose edges are
spacelike to a common extremal or generic trapped surface, then we expect that these diamonds
are independent, since they are contained in the domains of dependence of
the regions $\Sigma_R$ and $\Sigma_L$ that we expect to be independent -- see
Fig.~\ref{fig:diamonds2}.\footnote{Assuming that the edges are not
parametrically close to null separated, so that the splitting region is
parametrically small.}

\subsection{A Lorentzian analysis}\label{sec:lorentzian}
It  is illustrative to reach the same conclusion using a spacetime approach. Let
us now do this, at the same type generalizing to also allow planar or hyperbolic
symmetry, in addition to spherical. This will also let us show that arbitrarily large energy
densities near an extremal surface make arbitrarily small contributions to the
ADM mass.

Let $(N, g_{ab})$ be a $(d+1)$-dimensional spacetime with one of these three types of
symmetries. This means that $N$ can be
foliated by a two-parameter family of codimension$-2$ surfaces $\sigma$ that
have the intrinsic metric of a sphere, a
plane, or the hyperbolic plane (or quotients thereof). 
We can pick a double null
gauge
\begin{equation}
\begin{aligned}
    \dd s^2 = -2 e^{-f(x^+, x^-)}\dd x^+ \dd x^- + r(x^+, r^-)^2 \dd \Sigma_k^2,
\end{aligned}
\end{equation}
where $\dd \Sigma_k^2$ is the metric of the unit sphere, plane, and hyperbolic plane
for $k=-1, 0, 1$, respectively (or quotients thereof). For the leaves
of our symmetric foliation, given by constant $x^+,
x^-$, we now define the following function\footnote{$\mu$ is a generalization of the Lorentzian Hawking mass
\cite{MisSha64,Haw68,Hay98,BraHay06} to $d\neq 3$, in the special case of
spatial symmetries. Without symmetries, the proper definition of the Hawking
mass is unknown for $d\neq 3$.}
\begin{equation}
\begin{aligned}
    \mu(x^+, x^-) = kr^{d-2}  + \frac{ r^{d} }{ L^2 } - \frac{ 2\theta_+
    \theta_{-} }{ k_+ \cdot k_- (d-1)^2 },
\end{aligned}
\end{equation}
where $k_+^a, k_-^a$ are any two independent future-directed null normals to the
surface, and $\theta_+, \theta_-$ the corresponding null
expansions.
The
quantity on the RHS is covariantly defined, since the area radius $r$ can be
viewed as a coordinate independent scalar on spacetime.\footnote{This is not
strictly true in planar symmetry. In this case, there is an ambiguity in an
overall scaling of $r$, reflective of the fact that there is no canonical
conformal frame for the boundary of AdS, when the boundary is conformal to
Minkowski. In this case, there is also a scaling ambiguity in the mass. But this
ambiguity can be fixed in some particular spacetime.} $\mu$ is a spacetime analog of $\omega$, and it has some special properties. First, in AAdS or AF
spacetimes, it can be shown to reduce to the mass at spatial
infinity:\footnote{Provided matter fields fall off fast enough. The important thing is
that $\mu$ is a spacetime function with certain monotonicities that sometimes act as an obstruction to
independence.}
\begin{equation}\label{eq:mum}
\begin{aligned}
    M = \frac{ (d-1)\text{Vol}[\Sigma_k] }{ 16\pi G_N }\mu|_{r=\infty}.
\end{aligned}
\end{equation}
Second, using the Einstein equations, it was shown in \cite{Fol22} to satisfy
\begin{equation}\label{eq:mup}
\begin{aligned}
    \partial_{\pm}\mu = \frac{ 2e^{f}r^{d} }{ (d-1)^2 }(T_{+-}\theta_{\pm} -
    \theta_{\mp}T_{\pm\pm}).
\end{aligned}
\end{equation}
Let now $X^a$ be any spacelike vector pointing outwards ($X_a k_+^a \geq 0$), and assume the dominant
energy condition (DEC), which says that
\begin{equation}
\begin{aligned}
    T_{ab}U^a V^b \geq 0 \quad \forall \text{ timelike } U^a, V^a,
\end{aligned}
\end{equation}
and which implies the WEC. The DEC implies that $T_{\pm \pm}\geq 0, T_{+-}\geq 0$. Thus, in a ``normal'' region, where $\theta_+ \geq 0 , \theta_- \leq 0$,
$\mu$ is monotonically non-decreasing in any outwards spacelike direction:
\begin{equation}\label{eq:mumon1}
\begin{aligned}
    X^a \nabla_a\mu \geq 0 \qquad \text{when}\quad \theta_+ \geq 0, \quad
    \theta_{-}\leq 0.
\end{aligned}
\end{equation}
Next, in an ``anti-normal'' region, we have monotonicity in the inwards
direction instead:
\begin{equation}\label{eq:mumon2}
\begin{aligned}
    X^a \nabla_a \mu \leq 0 \qquad \text{when}\quad \theta_+ \leq 0, \quad
    \theta_{-}\geq 0.
\end{aligned}
\end{equation}
This is the spacetime analog of monotonicity of $\omega$, and it constitutes an obstruction to subregion independence
(if we break the DEC, this obstruction only becomes weaker). 
However, we see that if $\sigma$ is an extremal surface that separates a normal and an
anti-normal region, then the insertion of matter causes
$\mu$ to increase in opposite directions on opposite sides of $\sigma$, and so
we regain independence between the opposite sides. 
If we instead find ourselves in an (anti)trapped region of spacetime, $\mu$ is
no longer monotonic, and by the appropriate choice of turning on
either $T_{++}$ or $T_{--}$, we can push $\mu$ up or down as we are moving
in any fixed spacelike direction. 

We also see from \eqref{eq:mum} and
\eqref{eq:mup} that the contribution to the asymptotic mass is given by a
integral of $T_{++}\theta_-, T_{--}\theta_{+}, T_{+-}\theta_{-}, T_{+-}\theta_{+}$ weighted by
positive factors that are bounded in a neighbourhood of an extremal surface. Thus, arbitrarily large energy densities make
arbitrarily small contributions to the ADM mass, provided they are localized to
an extremal surface. Similarly, a marginally trapped surface (say,
$\theta_+=0$) can support modes
with very large $T_{--}$ at low cost to the ADM mass, provided we do not make
$T_{+-}$ large as well.

\subsection{No symmetries}\label{sec:IMC}
\begin{figure}
\centering
\includegraphics[width=0.55\textwidth]{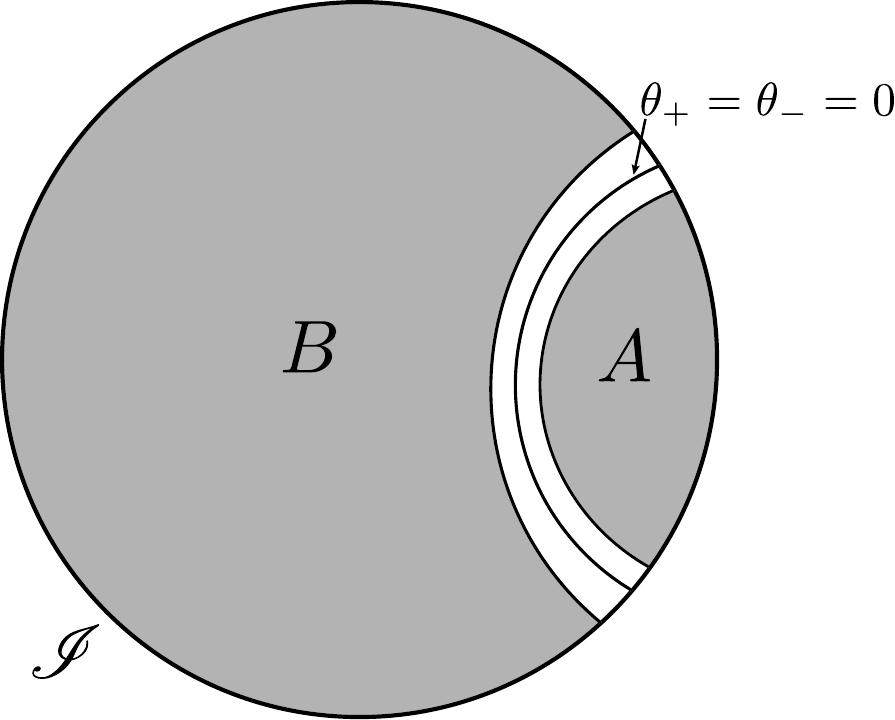}
    \caption{A timeslice $\Sigma$ of an asymptotically AdS spacetime containing
    two subregions $A$ and $B$ that are separated by a set containing an
    HRT surface anchored to the conformal boundary $\mathscr{I}$.}
\label{fig:ads}
\end{figure}

We now want to understand what happens if our spacetime does not have any
symmetries. Furthermore, even if our spacetime has symmetries, we might want to
consider perturbations or surfaces that break the symmetries of the spacetime. A
prototypical example of the latter occurs in AdS/CFT, where we might consider two
regions $A$ and $B$ separated by a region containing a boundary anchored extremal surface, 
as illustrated in Fig.~\ref{fig:ads}. 

To deal with the case of no symmetry, we need an appropriate generalization of
$\omega$. To the author's knowledge, the appropriate generalization is known
only in four spacetime dimensions, and so we will
only treat this case. However, based on expectations from AdS/CFT and the fact that the spherically
symmetric considerations work for all $d\geq 2$, we expect a similar story
holds in other dimensions, but we do not know what is the right tool to
use.

Let us thus assume four spacetime dimensions. If $(\Sigma, S)$ is some initial dataset with a spacelike two-dimensional surface $\sigma$, we define
\begin{equation}
\begin{aligned}
    \omega[\sigma] =
    \frac{ 1 }{ 16\pi
    }\sqrt{\frac{\text{Area}[\sigma]}{4\pi}}\int_{\sigma}\left[2\mathfrak{R}-H[\sigma]^2 +
    \frac{ 4 }{ L^2 } \right],
\end{aligned}
\end{equation}
where the integral is taken in the induced volume form on $\sigma$, and where
$\mathfrak{R}$ is the Ricci scalar for the induced metric on $\sigma$.
Consider
now a one-parameter family of surfaces $\sigma_{\tau}$, where increasing $\tau$
corresponds to flowing the surfaces along the vector field
\begin{equation}
\begin{aligned}
    v^a = \frac{ 1 }{ H[\sigma_{\tau}] }r^a_{\tau},
\end{aligned}
\end{equation}
with $r^a_{\tau}$ being a unit normal to $\Sigma$ in $\sigma$.
Thus, our one-parameter family of surfaces correspond to a flow where the
velocity of the flow at $p\in \sigma_\tau$ is set
by the inverse of the mean curvature of $\sigma_{\tau}$ at $p$. 
This flow is known as inverse mean curvature (IMC) flow,\footnote{In four
spacetime dimensions, a Lorentzian version
of IMC flow also exists \cite{Fra01,BraHay06,BraJau15}, but is less understood. It generalizes the
monotonicities \eqref{eq:mumon1} and \eqref{eq:mumon2} to cases with no symmetries.}
and it can be shown that on a slice with $K\indices{^a_a}=0$, assuming the WEC,
we have  \cite{Ger73}
\begin{align}
    \frac{ \dd }{ \dd \tau }\text{Area}[\sigma_{\tau}] &\geq 0,\label{eq:IMC1} \\
    \frac{ \dd }{ \dd \tau }\omega[\sigma_{\tau}] &\geq 0.\label{eq:IMC2}
\end{align}
This is the generalization of $\omega'(r)\geq 0$.
Minimal and maximal area surfaces are special locations where the flow terminates. 
See \cite{FisWis16} for a review of these facts for compact surfaces, and for
the proof that \eqref{eq:IMC1} and \eqref{eq:IMC2} remains true also for boundary
anchored surfaces in asymptotically AdS$_4$ spacetimes ($\omega$ is finite even
when $\sigma_{\tau}$ is boundary anchored \cite{FisWis16}). 

\begin{figure}
\centering
\includegraphics[width=0.9\textwidth]{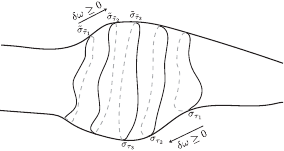}
    \caption{Two IMC flows $\sigma_{\tau}$ and $\tilde{\sigma}_{\tilde{\tau}}$
    accumulating at a maximal surface $\tilde{\sigma}_{\tilde{\tau_3}} = \sigma_{\tau_3}$.}
\label{fig:IMC}
\end{figure}

Consider a surface that is a minimal or maximal surface on a
$K\indices{^a_a}=0$ slice.  Then, as
illustrated in Fig.~\ref{fig:IMC}, IMC flow goes in opposite directions, so again we have a
mechanism to bypass the monotonicity of $\omega[\sigma_{\tau}]$. Positive energy density
can now likely screen positive energy density.  Of course, to show this rigorously is
challenging,\footnote{Note that
IMC flow can have singularities. In this case, a weak version of the flow can be
defined. In this case, when a surface becomes singular, it makes a jump to a surface with
larger area and Hawking mass. However, since we are trying to screen
perturbative excitations by adding matter near the minimal/maximal surface at
which we start the flow, should be able to screen the perturbation
before we reach a jump, since the jump-time is non-perturbative, so it will not
get radically affected by our perturbations.} but it would not be surprising if
it is true. It would fit perfectly with what we know about AdS/CFT, and 
it is plausible that the monotonicity of $\omega$ is the sole 
obstruction to subregion independence in gravity, since charges other than the
mass tend to not have a preferred sign. In fact, in various gluing results,
there is typically only a finite-dimensional space of obstructions to gluing,
corresponding to a set of charges that must match or have a certain relationship \cite{Cor00,CziRod22}. In four
spacetime dimensions, this is a set of 10 charges, and the mass is the only one
with a preferred sign.\footnote{The other charges corresponds to linear
momentum, angular momentum, and center of mass.} Thus, we conjecture the following:
\begin{conj*}
    Let $(\Sigma, S)$ be a smooth $d$-dimensional initial dataset in Einstein gravity minimally coupled to matter, with
    $K\indices{_a^a}=0$, and $d\geq 3$. Let $A,
    B$ be two closed disjoint subregions such that $\Sigma-A-B$ contains a locally minimal or
    maximal surface that is homologous to $\partial A$ and $\partial B$. For any
    sufficiently small deformations $\delta S|_A$ and $\delta S|_B$, there exists a small extension $\delta S$ of $\delta S|_A \cup \delta
    S|_{B}$.
\end{conj*}
We are deliberately vague about what we mean with small here. We might either consider
formal perturbative solutions, or perhaps better, we could try to require that with an 
appropriate choice of $C^k$ or Sobolev norm defined by the background metric, 
a small extension always exists once the norm of $\delta
S|_A \cup \delta S|_{B}$ is sufficiently small. Note that the above might also
be true for $d=2$ if matter is included, but for $d=2$ vacuum gravity, we saw
that we were forced to change the topology of $C$ to match perturbations made in
$A$, $B$, so there was no sense in which the perturbation was small.

\begin{figure}
\centering
\includegraphics[width=0.4\textwidth]{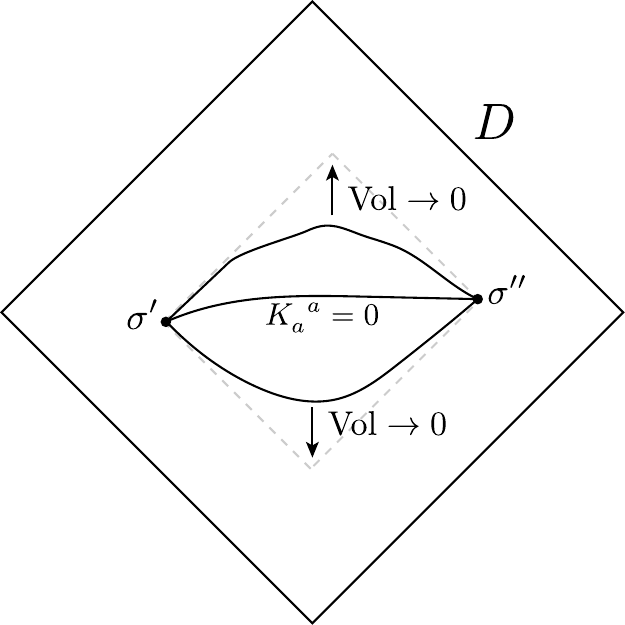}
    \caption{A small domain of dependence $D$. For any two spacelike separated
    surfaces $\sigma', \sigma''$, we expect there to be a maximal volume slice
    $\Sigma$ with $\partial \Sigma = \sigma' \cup \sigma''$.}
\label{fig:maxvoldiamond}
\end{figure}

Next, we again really want to talk about surfaces that are trapped or extremal,
rather that maximal or minimal on a $K\indices{_a^a}=0$ slice. 
By the same logic as in the spherically symmetric discussion, we thus want that every
extremal and generic trapped surface $\sigma$ is minimal or maximal on some
slice $\Sigma$ that satisfies $K\indices{_a^a}=0$ locally in a neighbourhood of
$\sigma$. While we will not attempt a proof, this  appears very likely to be
true. Consider a region of spacetime $D$ corresponding to a domain of dependence
containing $\sigma$. Let us shrink $D$ enough so that its closure is compact,
so that the null generators of its boundary do not terminate on singularities or
future infinity. Then for any two spacelike separated surfaces $\sigma', \sigma''$
in $D$,
there should exist a maximal volume hypersurface $\Sigma$ bounded by $\sigma', \sigma''$.
See Fig.~\ref{fig:maxvoldiamond}. This makes sense, since if $\Sigma$ is deformed towards the future
or past null congruences fired from $\sigma', \sigma''$, its volume goes to
zero. Provided we pick $D$ small enough so that it does not contain a portion of a
future/past infinity where volume of space could be forever expanding, like
future infinity of de Sitter, we should not be able to make the volume of $\Sigma$ arbitrary
large. Thus, there ought to be a maximal volume slice bounded by
$\sigma', \sigma''$. This surface is necessarily a $K\indices{_a^a}=0$ slice,
and so in $D$ we ought to have a $K\indices{_a^a}=0$ slice for every choice of two spacelike
separated surfaces $\sigma', \sigma''$. Allowing these surfaces to vary, this likely provides more
than enough freedom for finding such a slice that contains $\sigma$, again suggesting that the setup
described by Fig.~\ref{fig:diamonds2} is true without spherical symmetry as well.

\section{de Sitter rigidity and area bounds}\label{sec:dS}
In our discussion of pure AdS and Minkowski, we saw that these spacetimes were very
rigid. No deformation in the interior of the spacetime can be made without it
altering the geometry at infinity. Can some form of rigidity statement be made for dS?  
A natural question to ask is the following: is it possible to deform the initial
data in one static patch without altering the other? At the level of spherical
symmetry, we saw that the answer was no at first order in perturbation theory.
We will now prove a non-linear version of this statement. However, as we discuss
below, once we break spherical symmetry, rigidity appears to no longer hold,
unlike the case of Minkowski and AdS. This likely gives rise to large class
of spacetimes that look identical to dS in a single static patch.

\begin{figure}
\centering
\includegraphics[width=0.7\textwidth]{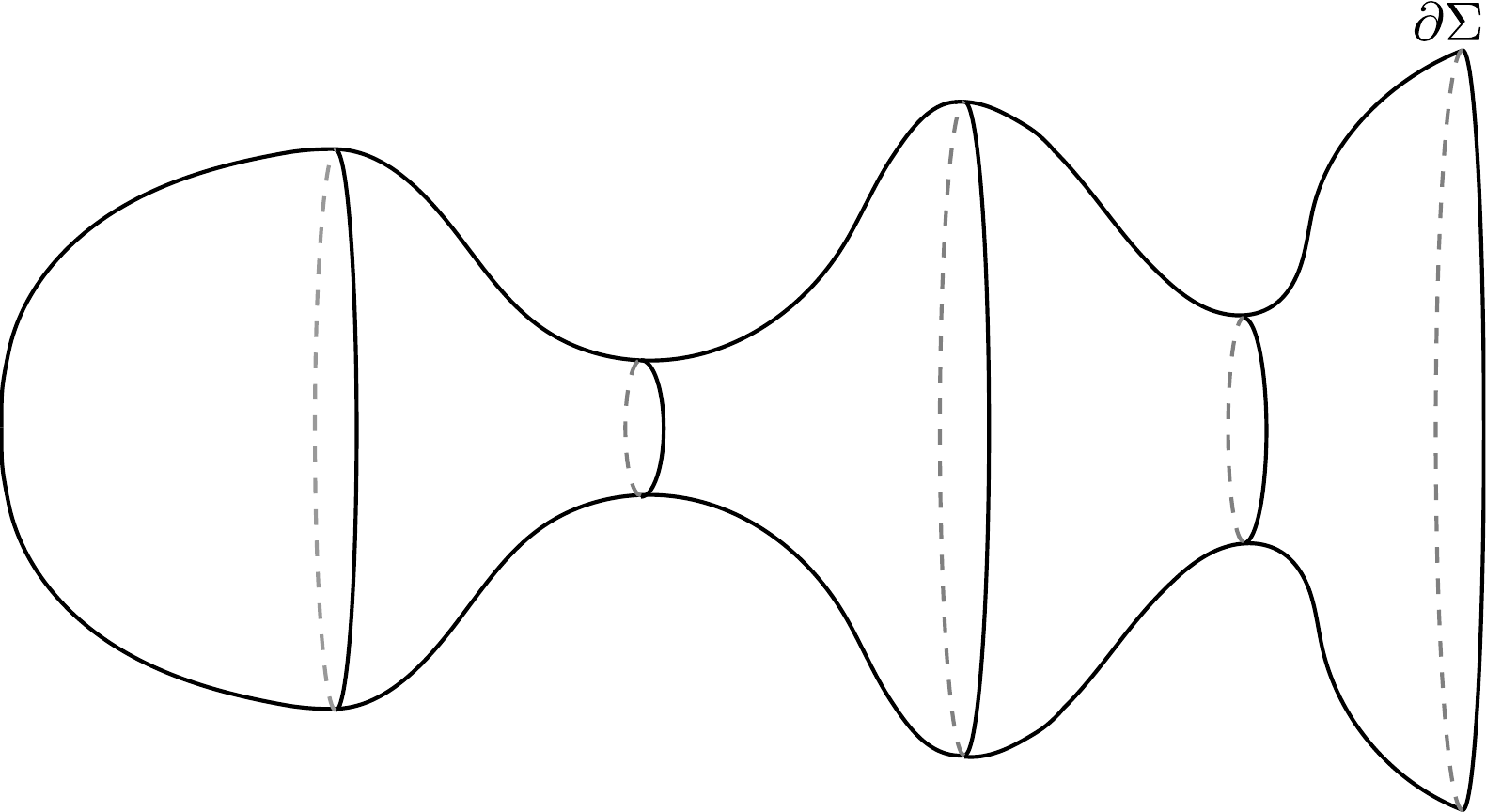}
    \caption{A spherically symmetric spatial manifold with a smooth $r=0$ and a
    locally maximal boundary. When $K\indices{_a^a}=0$, the WEC implies that all
    spheres satisfy $r\leq L_{\rm dS}$}
\label{fig:dSslice}
\end{figure}

We prove the following\footnote{The author is grateful to Matt Headrick for proposing that Nariai
might set a lower area bound, and for extensive discussions that lead directly to this result.}
\begin{thm}\label{thm:1}
    Let $(\Sigma, S)$ be a regular spherically symmetric initial dataset for
    the Einstein equations with positive cosmological constant,
    $K\indices{_a^a} = 0$, and satisfying the
    WEC. If $\Sigma$ has the topology of a
    hemisphere (i.e. a ball) of dimension $d\geq 2$, then every sphere $\sigma$
    in $\Sigma$ has an area radius $r$ satisfying
    \begin{equation}\label{eq:rupL}
    \begin{aligned}
    r \leq L_{\rm dS}.
    \end{aligned}
    \end{equation}
    Equality is achieved if and only if $K_{ab}=0$, with $h_{ab}$ the metric of
    a round sphere. Next, if $\sigma_*$ is a
    locally maximal sphere of radius $r_*$, then
    \begin{equation}\label{eq:rlower}
    \begin{aligned}
      r_* \geq \sqrt{\frac{ d-2 }{ d }} L_{\rm dS} = r_{\rm Nariai}.
    \end{aligned}
    \end{equation}
\end{thm}
\begin{figure}
\centering
\includegraphics[width=0.6\textwidth]{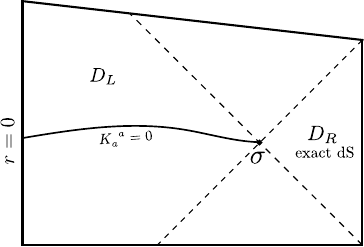}
    \caption{A hypothetical spherically symmetric spacetime that is ruled out by
    Theorem~\ref{thm:1} when the WEC holds.}
\label{fig:impossible}
\end{figure}
Before the proof, let us make a few comments. First, at
leading order in perturbation theory around pure dS, the upper bound \eqref{eq:rupL} is just a
special case of the first law \cite{GibHaw77b}. Next, for $d=3$, the upper bound
\eqref{eq:rupL}, which is equivalent to $A(\sigma) \leq 4\pi L_{\rm dS}^2$, was proven
without spherical symmetry in \cite{ShiNak93}, given certain other assumptions
(see also \cite{ShiIzu22}). Second, extremal
surfaces are always stationary on any slice, so for an extremal surface $X$ in a
spherical asymptotically dS spacetime with a simply connected Cauchy slice, we
expect the area bounds \eqref{eq:rupL}, \eqref{eq:rlower} to apply to $X$,
provided $X$ is a cosmological horizon type surface -- i.e. it is maximal rather
than minimal/a saddle on $\Sigma$. Third, we are not aware if a bound like
\eqref{eq:rlower}, which lower bounds the area of cosmological horizon-type
surfaces in terms of the event horizon of the Nariai black hole, has been shown
before (beyond the dS-Schwarzschild family).  Finally, the result shows that a spacetime
with the following properties cannot exist when the WEC holds: (1) $\sigma$ is a
sphere that splits a Cauchy slice in two, (2) the wedge $D_R$ of points
right-spacelike to $\sigma$ looks identically like a static patch of de Sitter,
and (3) the wedge of $D_L$ points left-spacelike to $\sigma$ has a Cauchy slice
$\Sigma$ with $K\indices{^a_a}=0$. See Fig.~\ref{fig:impossible}. Now to the
proof:
\begin{proof}
    By the fact that $\Sigma$ is spacelike, we have
    \begin{equation}
    \begin{aligned}
        1 - \frac{ r^2 }{ L_{\rm dS}^2 } - \frac{ \omega(r) }{ r^{d-2} } \geq 0,
    \end{aligned}
    \end{equation}
    as discussed in Sec.~\ref{sec:setup}. Thus, if we can show that
    $\omega(r)\geq0$ on every patch of $\Sigma$, 
    then we get that $r \leq L_{\rm dS}$. Let us now show this.
    The Hamiltonian constraint together with $K\indices{_a^a}=0$ implies that $\omega'(r)\geq0$. 
    Furthermore, by
    assumption, $\Sigma$ contains the point $r=0$, where we must have
    $\omega(0)=0$. Hence, integrating
    \eqref{eq:fullConstraint} outwards, we find that $\omega \geq 0$ on the patch
    containing $r=0$. If one patch covers $\Sigma$, we are done. Thus, assume
    that $\Sigma$ is covered by multiple patches. Then the $r=0$ patch is
    separated from the next patch by either a maximal surface or a saddle -- see 
    Fig.~\ref{fig:dSslice}.
    If it is a saddle, it does not alter the monotonicity properties of
    $\omega$ as we move from $r=0$ towards $\partial \Sigma$, so $\omega$
    remains positive on the next patch. Thus, assume we have a maximal surface
    instead. Consider first integrating $\omega$ from $r=0$ to the first maximal surface,
    with radius $r_{\rm max}$. 
    We already showed that $\omega \geq 0$ there, so we find that
    $r_{\text{max}} \leq L_{\rm dS}$. Transitioning to the next patch, $r$ and
    $\omega$ must
    now be decreasing as we move towards $\partial \Sigma$. Either we hit
    $\partial \Sigma$ and we are done since then $r|_{\partial \Sigma}<r_{\rm
    max}\leq L_{\rm dS}$, or we hit a minimal surface with
    radius $r=r_{\rm min} < r_{\rm max} \leq L_{\rm dS}$ (again, we can hit a
    saddle, but nothing interesting happens at these). By stationarity we get
    \begin{equation}
    \begin{aligned}
        \frac{ \omega(r_{\rm min}) }{ r_{\rm min}^{d-1}} =  1 - \frac{ r_{\rm
        min}^2 }{ L_{\rm dS}^2 } > 1 - \frac{ r_{\rm
        max }^2 }{ L_{\rm dS}^2 } \geq 0.
    \end{aligned}
    \end{equation}
    Repeating this exact argument as we move past any number of minimal,
    maximal, or saddle surfaces, we find that $\omega \geq 0$ everywhere,
    showing that $r\leq L_{\rm dS}$. Next, we only find a sphere with $r=L_{\rm
    dS}$ if $\Sigma$ a subset of exact de Sitter.  This is seen by the fact that chain
    of potential equalities above is broken if we encounter matter anywhere.  This proves rigidity.

    To prove the lower bound, note that approaching a stationary surface
    implies that $1/B(r)$ is approaching $0$ from a positive value. The
    surface being maximal with radius $r=r_*$ implies that we are approaching $r_*$ from a
    smaller value of $r$. Thus, $\frac{ \dd }{ \dd r }B^{-1}|_{r=r_*}\leq 0$,
    giving that
    \begin{equation}
    \begin{aligned}
        0 &\geq r_* \frac{ \dd }{ \dd r }\left[1-\frac{ r^2 }{ L_{\rm dS}^2 } - \frac{
            \omega(r) }{ r^{d-2} }\right]\Big|_{r=r_*} \\
        &= - \frac{ 2r _*^2 }{ L_{\rm dS}^2 } - \frac{ \omega'(r_*) }{ r_*^{d-3}
        }+(d-2)\frac{ \omega(r_*) }{ r_*^{d-2} }\\
        &= -\frac{ dr_*^2 }{ L_{\rm dS}^2 } +d-2 - \frac{ \omega'(r_*) }{ r_*^{d-3}
        }.
    \end{aligned}
    \end{equation}
    Using that $\omega'\geq 0$, this then gives the lower bound on $r_*$.
\end{proof}

Let us finally discuss the case without spherical symmetry, which turns out
to be interesting. It would be tempting to
conjecture the following: if $\Sigma$ is a $K\indices{_a^a}=0$ slice where
$\partial \Sigma$ is an extremal surface, then $A[\partial \Sigma]\leq
A[\partial \Sigma_{\rm dS}]$, with equality only in the case of pure dS
(assuming the WEC). While the area inequality might hold, the rigidity part, i.e.
equality only in pure dS, is overwhelmingly likely to be false for $d\geq 3$.
The rigidity part is closely related to a former conjecture by Min-Oo
\cite{MinOo02}, who conjectured the following: assume $h_{ab}$ is
a metric on a hemisphere $\Sigma$ with (1) a Ricci scalar lower bound
$\mathcal{R}\geq d(d-1)L_{\rm dS}^{-2}$,\footnote{This is the lower bound one would get from the constraints, assuming the WEC and $K\indices{_a^a}=0$.}
and (2) with the intrinsic and extrinsic geometry of $\partial
\Sigma$ matching that of the boundary of a round hemisphere of radius $L_{\rm dS}$. Then $\Sigma$
must be the round hemisphere.\footnote{With hemisphere, we mean that $\partial
\Sigma$ is totally geodesic in $\Sigma$ -- it is really a half sphere, not just a cap.}
While according to \cite{BreMar10} this conjecture was proven in \cite{Top59} for $d=2$,
after being open for
16 years, Min-Oo's conjecture was disproven in \cite{BreMar10} for all $d\geq
3$. They showed that there are 
Riemannian metrics distinct from the canonical round sphere metric that
nevertheless looks identical to it in a neighbourhood around $\partial \Sigma$,
for any $d \geq 3$. Several properties of these solutions are known. They can
be constructed to have arbitrarily large volume \cite{CorEic13},\footnote{None of the
solutions with volume greater than that of the static patch lie perturbatively
close to the static patch, since the results of \cite{PenLue11} imply that
WEC-respecting metrics $C^2$-close to the static patch must have smaller volume.}
they always
have a minimal surface \cite{MarNev12}, and they can have a wide range of
topologies \cite{CorEic13, Swe23}. Furthermore, for $h_{ab}$ sufficiently close to pure
dS, the metric must disagree with the round metric somewhere in the band \cite{BreMar11}
\begin{equation}
\begin{aligned}
    \sqrt{\frac{ d-1 }{ d+3 }}L_{\rm dS} < r < L_{\rm dS},
\end{aligned}
\end{equation}
so these deformations cannot localize in a tiny cap. 
If any of these metrics can be realized as a solution of the constraints,
either with some choice of $K_{ab}\neq 0$ and/or with some choice of matter, then the rigidity part of the upper bound in
Theorem~\ref{thm:1} cannot be true without spherical symmetry (when
$d>2$).  Thus, in four spacetime dimensions and higher, when we break symmetries
there might exist a large flexibility in changing the initial data in only
one static patch.  The behavior of these solutions are reminiscent of another
example we have already discussed: black holes in GR with a negative
cosmological constant and three spacetime dimensions. In this case there is an
infinite number of one-sided black holes that look identical to the BTZ black
hole in the exterior(s). However, they all break spherical symmetry in
the interior, and none are close to a spherically symmetric metric. Like black holes, the de Sitter metrics discussed above also have
minimal surfaces, so it seems likely that they will be black
holes as well.  It would be interesting if these could be used to give a
semiclassical counting interpretation of the area of the cosmological horizon
using the Euclidean path integral approach of \cite{PenShe19,BalLaw22}.

\section{Discussion}\label{sec:discussion}
In this work, we have given a simple natural definition of subregion
independence in classical gravity. We argued that extremal surfaces, generic
trapped surfaces, and background distributions of matter are structures that
enable subregion independence, i.e. independent initial data perturbations.
For extremal surfaces $X$, we saw that excitations added on the opposite side of
$X$ contribute with opposite sign to any given asymptotic mass. This enables
positive energy densities to terminate the gravitational fields sourced by positive
energy densities on the other side of $X$, providing a simple physical
picture for why an extremal surface is a good location to separate independent
subregions.

We now discuss some further implications of our perspective.

\subsection{The semiclassical case and quantum extremal surfaces}
Consider studying semiclassical gravity, where we couple the expectation value
of the stress tensor of quantum matter to the classical Einstein equation. In
this case, the constraint equations are unchanged, so our analysis is still
relevant, albeit incomplete. The matter source is now more exotic, since 
classical energy conditions are violated. However, in our case, the
classical energy conditions were an obstruction to independence, so we expect
that this particular effect pushes subregions to a greater tendency for independence.
Nevertheless, while quantum fields typically do not have local energy conditions, they
often have global ones, so we do not expect that the problem is trivial -- the
Hawking mass in a normal region still ought to have certain monotonicity properties over
sufficiently large distances. 

Next, while our analysis does not directly say anything about general
QESes \cite{EngWal14}, our results are informative in many particular cases. 
We might for example find that QESes themselves are classically (anti)trapped,
so that we still avoid monotonicity of the Hawking
mass, boding well for the perturbative independence of the complement wedges of the QES.
Next, if a QES is a Planckian distance away from a classical extremal 
or (anti)trapped surface, we can tell a similar story, since we always want to
draw some splitting region around the QES, which then contain a
classically trapped or extremal surface. However, we leave a more careful study
of QESs to the future. 

\subsection{Islands in massless gravity}
The new QESs discovered after the Page time \cite{AlmEng19,Pen19}, giving rise to the island
phenomenon, are at the heart of recent breakthroughs on the information paradox \cite{AlmEng19,Pen19,PenShe19,AlmHar19}. Islands are regions of
spacetime that furnish entanglement wedges for the Hawking radiation after the
Page time. They have the special property that they are compact, not reaching
any asymptotic boundaries. Since the island is supposed to be encoded in the
radiation, acting with operators in the island should correspond to acting with
operators on the radiation. 

In \cite{GenKar21} it was argued that this
behavior is inconsistent with having massless gravity. An important part
of their argument was the following claim: any excitation localized to the island must have
compact support, and by the Heisenberg uncertainty principle it should have
finite energy, thus altering the ADM energy. However, as should be clear by now,
non-linearities of the Einstein equations make this is false at the
classical level. While turning on excitations
 adds positive energy density locally, this does not imply that the ADM energy is
changed. It is true for perturbations around Minkowski space or AdS, but these
examples turn out to not be good analogies for the general case.
As discussed in Sec.~\ref{sec:spheresym}, in trapped regions of spacetime, 
ingoing and outgoing modes can be added so as to affect the 
the ADM mass with either sign. In the case of the evaporating black holes studied in \cite{AlmEng19,Pen19}, 
the island is in a classically trapped region of spacetime, 
so there is no obstruction to implementing large classes of localized
perturbations of the constraints with no influence on the geometry outside the
island. 

Another part of the argument in \cite{GenKar21} was the closely related claim that 
diffeomorphism invariant operators that probe local physics in the island must be dressed to the boundary, and so they must have
non-zero commutators with operators in the complement entanglement wedge,
leading to a contradiction with subregion-subregion duality. However, at the
level of perturbative quantization in $\sqrt{G_N}$ around some
background, there does exist localized operators that are not dressed to the
boundary \cite{Kom58,BerKom60,GidMar05,Tam11,Mar15,Kha15,GoeHoe22}. Operators like this were constructed in the CFT by
\cite{BahBel22,BahBel23}, and they argued this relieves the tension between
islands and massless gravity. We agree, and will add a few additional remarks.\footnote{See also
\cite{Kri20, GhoKri21}, which finds the appearance of islands in braneworlds with massless gravity.}

It was suggested in \cite{BahBel22,BahBel23} that one might need to
dress operators to the Hawking radiation. While perhaps possible, this can be
avoided for the evaporating black holes of \cite{AlmEng19,Pen19}.
To begin with, let $\Phi_0$ be some scalar field that is part of the background we are considering, and let
$\phi$ be a dynamical quantum field.\footnote{$\Phi_0$ could also be 
quantum field with a non-zero VEV.} Then
\begin{equation}\label{eq:Oalpha}
\begin{aligned}
    O(\alpha) = \int \dd^{d+1}x \sqrt{-g} \delta\left(\Phi_0(x)- \alpha \right)
    \phi(x)
\end{aligned}
\end{equation}
is a one-paramater family of diffeomorphism invariant observables. In the quantum
case, we should smear this over a window of $\alpha$-values to get a proper operator, but as
long as the gradient of $\Phi_0$ is not extremely small (i.e. scaling with $G_N$ to a
positive power), this gives an
approximately localized operator on appropriate backgrounds. The existence of
single-integral observables like \eqref{eq:Oalpha} was
acknowledged in \cite{GenKar21}. However, they argued that single-integral observables cannot probe large parts of the
island. This is based on the fact that once we reach the
Page time, the black hole interior volume (on some nice slice) has grown
approximately linearly for a time of
order $t\sim \mathcal{O}(G_N^{-1})$, so unless the black hole has been
constantly fed matter, there are large regions in the black hole where there is
no matter, or where the matter is extremely dilute.
So if we try to get the delta function in $O(\alpha)$ to ``click'' somewhere in the dilute
region, once we smear $\alpha$ a tiny bit, the operator strongly delocalizes.
There are however other better operators that do not have this problem. First, note that the matter
that collapsed to form the black hole never dilutes, and it is present in
the island.\footnote{Alternatively, if we instead evaporate a past-eternal black
hole in AdS, the
same goes through for the shell of matter that falls into the black hole due to
the sudden coupling of the CFT to the reservoir where we dump the Hawking
radiation.} So for some windows of $\alpha$, we could get operators
localized inside the matter distribution that formed the original black hole. 
But we can do better. Let us now consider forming the
following three-parameter family of diffeomorphism invariant
observables\footnote{Here $g_{ab}, \nabla_a$ are the full metric and
connections, so these should be expanded in a perturbative series. Similar for
$X_{\alpha,\beta,\gamma}$.}
\begin{figure}
\centering
\includegraphics[width=0.8\textwidth]{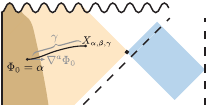}
    \caption{A diffeomorphism invariant operator localized to the island, dressed to matter that formed
    the black hole, and which probes the ``barren'' part of the island without
    becoming highly delocalized as $G_N \rightarrow 0$.}
\label{fig:island}
\end{figure}
\begin{equation}\label{eq:Oabc}
\begin{aligned}
    O(\alpha,\beta,\gamma)  = \int \dd^{d+1}x \sqrt{-g}
    \delta\left(\Phi_0(x) - \alpha\right)
    \delta\left(\nabla_a \Phi_0(x)\nabla^a \Phi_0(x) -
    \beta\right)\phi(X_{\alpha,\beta,\gamma}(x))
\end{aligned}
\end{equation}
where the $X_{\alpha,\beta,\gamma}(x)$ is the point obtained
when firing a geodesic from $x$ along the direction of $t^a = \nabla^a \Phi_0$ and
following it for a proper distance/time $\gamma$. See Fig.~\ref{fig:island}. Since we get to fix the sign
of $\beta$, we also get to fix the signature of $t^a$, so we know whether we are
using a spacelike or timelike geodesic -- i.e. the observable is well defined. To get an observable with the potential
to be promoted to a proper operator when quantizing, we could
smear over a small window of $\alpha, \beta, \gamma$. If we pick
$\alpha$-window
appropriately, then the $x$-integral can localize within the concentrated matter that formed
the black hole. Next, by increasing $\gamma$ up to values of order
$\mathcal{O}(1/G_N)$, we can reach into the ``barren'' matterless region of the
island.\footnote{Note that
setting $\gamma\sim \mathcal{O}(1/G_N)$ raises questions as to whether the operator
is well controlled in perturbation theory. However, the same issue would arise
for any boundary-dressed operator probing sufficiently deep into the interior around the
Page time. Thus, the issue of the consistency of islands in massless gravity does
not appear to have any bearing on this subtlety. At the very least, this
argument against the existence of islands would be an equally good argument
against the existence of the interior.} 
We could also reach this region by dressing to matter that falls into the black
hole around the Page time, if such matter exist.

All in all, perturbation theory in massless gravity appears to be consistent
with compact entanglement wedges. Of
course, once we worry about exponential corrections in $G_N$, things are much
more subtle.\footnote{See \cite{Jaf16} for an illuminating discussion of the
challenges of defining non-perturbatively diffeomorphism invariant observables.}
But once we consider exponential corrections, we should worry about what right
we have to talk about concepts like spacetime regions or entanglement wedges,
and it is hard to draw strong conclusions either way. But we see no clear sign
of inconsistencies between islands and massless gravity at the perturbative level.

\subsection{Gravitational splitting and local algebras}
In this paper, we have argued that classical gravity has a sort of
perturbative split property, provided the splitting region is sufficiently
generic -- i.e. it contains an
extremal surface, a non-perturbative amount of matter, or a generic (anti)trapped surface. It
would be very interesting to study the perturbative WdW equation to see whether
this remains true in the quantum case. 
Specifically, working around a
sufficiently generic background, can we choose the WdW wavefunctional on generic
separated spacelike subregions separated by a finite gap independently?
\cite{ChoGod21} showed that this is not true around pure AdS, but there are many hints that  
 this story changes on other backgrounds.
If we indeed can choose the WdW wavefunctional independently on different
generic regions (up to the local constraints on these regions), this suggest that
 non-trivial perturbative algebras of localized observables exist for 
 generic compact regions. This would be
 good news, given the recent interesting developments on algebras and their
 entropies in gravity
 \cite{LeuLiu21,Wit21a,LeuLiu22,ChaPen22,ChaLon22,JenSor23,PenWit23,AliSha23,Kol23,KliLei23,Ges23,KudLeu23,ColDon23,Wit23}. 

Note that we have not directly discussed the relation between independence and
localized observables, which is strictly speaking more directly tied to algebras that the states
themselves. We now turn to this.

\subsection{From independence to localized observables}\label{sec:obs}
Our notion of independence deals with the structure of the phase space
$\mathcal{P}$ of GR in a sufficiently small neighbourhood $U \subset
\mathcal{P}$ around a point $\Psi \in \mathcal{P}$, which corresponds to our
background.  It would be interesting to understand whether the independence of
$A, B \subset \Sigma$ means that there exist observables $f_{AC}, f_{BC}: U \mapsto
\mathbb{R}$ that are sensitive only to the dynamical fields in the domains of
dependence of $A\cup C$ and $B \cup C$, respectively. For example, if $B$ is a
collar around spatial infinity, $f_{AC}$ would correspond to a (perturbatively)
diffeomorphism invariant observable that is not dressed to the boundary, and it
would Poisson-commute with observables localized to $B$ \cite{Mar15}.  It would be natural if
independence implied that such functions exist. If they do, they are candidates
for localized operators upon quantization. Let us outline a non-rigorous
argument that localized observables exist. We do not claim that this completely
settles the issue -- there might be devils in the details.

Assume that there exist some gauge fixing procedure, so that the coordinate
values of $\partial A, \partial B$ are fixed, and so that every point $\delta \Psi \in U$
has a unique value of $\delta S|_{A}$, $\delta S|_{B}$ and $\delta S|_{C}$.
These three collections of classical field values on $A, B, C$ are generally
dependent on each other. However, if $A$ and $B$ are
independent, then $\delta S|_{A}$ and $\delta S|_{B}$ can be specified
independently. So they provide valid coordinates on $U$, and we can write a
coordinate representation 
\begin{equation}\label{eq:psiCoord}
\begin{aligned}
    \delta \Psi = \begin{pmatrix}
\delta S|_A \\
\delta S|_B \\
        \widehat{\delta S}|_C
\end{pmatrix},
\end{aligned}
\end{equation}
where $\widehat{\delta S}_C$ is a collection of coordinates parametrizing the remaining freedom
in $C$ independent of the degrees of freedom in $A$ and $B$. Clearly, by the
constraints, some of the degrees of freedom in $C$ are dependent on the degrees
of freedom in $A\cup B$, so  $\widehat{\delta S}|_C$ must be a strict subset of $\delta
S|_C$.

Now, since $\delta S|_A, \delta S|_B$ can be
used as coordinates,  the phase space coordinate functions
in the $A$-- and $B$--slots of \eqref{eq:psiCoord} then seem like candidates for $f_{AC}, f_{BC}$. It
might be tempting to say that these functions are sensitive to the dynamical
fields in just $A$ or just $B$, since these functions always return information
about the dynamical fields in $A$ or $B$. However, this is not true. A choice of $\delta S|_C$,
influences the possible values of $\delta S|_{A}$, and thus the possible output of 
the coordinate functions of the $A$-slot. Disregarding diffeomorphism invariance
for a second, an analogy in a theory with a
dynamical field $\phi$ and two points $x_A \in A, x_C \in C$ would be a function
like $f_{AC} = \phi(x_A)\theta(\phi(x_C))$, where
$\theta$ is the heaviside step function. 

Note that if $A$ and $B$ were dependent, this construction would not work. 
The possible values of $\delta S|_A$ would depend on the state on $B$, so any
function of $\delta \Psi$ that returned $\delta S|_A$ could not be sensitive
only to the fields in $A$, since the possible value of the fields there depends
on the fields in $B$.

\subsection{The Python's lunch}
\begin{figure}
\centering
\includegraphics[width=0.7\textwidth]{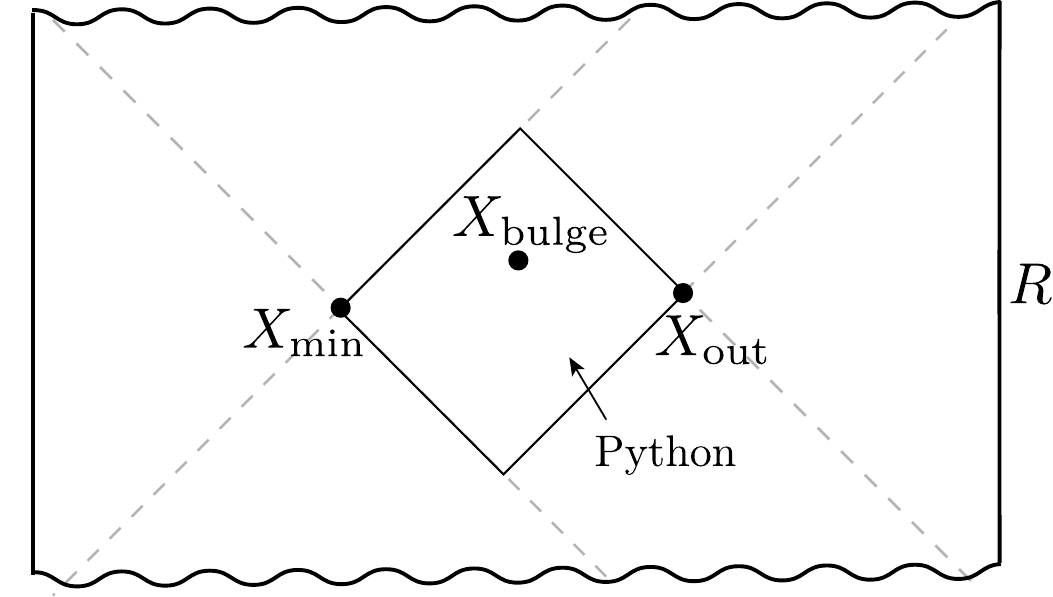}
    \caption{A spacetime with a Python's lunch for a complete boundary component
    $R$. The wedge to the left and right of
    $X_{\rm min}$ are the entanglement wedges of the left and the right CFT,
    respectively.}
\label{fig:python}
\end{figure}
Consider an asymptotically AdS spacetime, and let $R$ be a spatial subregion of the
conformal boundary -- or perhaps a complete connected component. Next, assume that $R$ has three extremal
surfaces homologous to it: $X_{\rm min}$, $X_{\rm out}$ and $X_{\rm bulge}$, as
shown in Fig.~\ref{fig:python} in the case where $R$ is a complete boundary
component. $X_{\rm min}$ is the HRT surface, which defines the
entanglement wedge. $X_{\rm out}$ is an extremal surface forming a candidate
HRT surface, but it has area greater than $X_{\rm min}$. $X_{\rm bulge}$ lies in
between the two, has greater area than either, and it is not minimal on any
spatial slice. These three surfaces together form a structure known as a
``Python's lunch'' -- see Fig.~\ref{fig:python}. It was conjectured in \cite{BroGha19} that the
computational complexity required to reconstruct operators in the Python's lunch
from the CFT scales as
\begin{equation}
\begin{aligned}
    \propto \exp\left( \frac{ \text{Area}[X_{\rm bulge}] -
    \text{Area}[X_{\rm out}] }{ 8 G_N  }\right).
\end{aligned}
\end{equation}
In the semiclassical case, we replace HRT with QES and
$\text{Area}/4G_N$ with the generalized entropy. Thus, the reconstruction of
operators in the lunch from operators near the boundary is non-perturbative in $G_N$, and should not be
achievable when just accessing the perturbative bulk description (including gravitons). Further evidence
for this was found in \cite{EngPen21}, where they showed that classical boundary sources, together with repeated forwards and background time
evolution (i.e. time-folds), cannot be used to expose the lunch region and
make its data causally accessible from the boundary. 

The findings in this paper play nicely with the Python's conjecture. The bulge
provides a surface that can be dressed across, giving it a functional role.  It enables perturbations with
spatial compact support around the bulge, never disturbing the complement wedges
to the lunch, including the exterior of $X_{\rm
out}$, which is the part of the entanglement wedge that is believed to be simple
to reconstruct. Furthermore, at least in spherical symmetry, bulge surfaces are supported
by matter, which provides additional background structures to anchor
diffeomorphism invariant with respect observables to. So it is quite reasonable that the perturbatively quantized theory
supports a non-trivial localized algebra of observables inside the Python that
commutes with observables in the simple wedge, and thus observables near the
boundary, to all orders in perturbation theory in $G_N$ (see \cite{EngLiu23} for a recent
discussion of this algebra).

\subsection{Entanglement wedges of gravitating
regions}
In \cite{BouPen22, BouPen23} they proposed how to define entanglement wedges of
general gravitational subregions $A$. More precisely, they defined a
min-entanglement wedge $e_{\rm min}(A)$ and a max-entanglement wedge $e_{\rm
max}(A)$. $e_{\rm max}(A)$ is proposed to be the region where the semiclassical
description can be fully reconstructed from operators associated to $A$, while
$e_{\rm min}(A)$ is the complement of the largest wedge about which nothing can
be learned from operators associated to $A$. Consider now two regions $A, B$
such that $A \subsetneq e_{\rm min}(B)^c$ and $B \subsetneq e_{\rm min}(A)^c$, where
$c$ denotes the complement, and where $X\subsetneq Y$ is used to indicate
$\partial X \cap \partial Y = \varnothing$. Let's call this EW-independence of
$A$ and $B$. From the above proposal, we should then have that 
semiclassical operators in $A$ and $B$ can be implemented completely
independently. It is then natural to conjecture that 
\begin{equation}\label{eq:EWindep}
\begin{aligned}
    \text{EW-independence of } A \text{ and } B \Rightarrow \text{classical
    independence of } A \text{ and } B,
\end{aligned}
\end{equation}
where classical independence is the notion studied in this paper.
Note however that the operators associated to $A$ that can reconstruct the relevant wedges are part of some yet-to-be-understood
quantum mechanical system associated to $A$, rather than the semiclassical
degrees of freedom of $A$. As a consequence, we do not expect the converse of
\eqref{eq:EWindep} to hold. Even if
$A$ and $B$ are classically independent regions, the fundamental degrees of freedom
associated to $A$ and $B$ that are capable of entanglement wedge reconstruction
might not be independent. It is interesting to consider a case of a region $A$ and
a region $B \subsetneq e_{\rm max}(A)-A$, with $A$ and $B$ classically independent. 
In this case $A$ can presumably still reconstruct $B$ with access to potentially
non-semiclassical quantum effects. Could this gap signify 
that it is non-perturbatively difficult to reconstruct $B$ from $A$?

\section*{Acknowledgements}
It is a pleasure to thank Chris Akers, Raphael Bousso, Roberto Emparan, Netta Engelhardt, Daniel
Harlow, Matt Headrick, Philipp Hoehn, Adam Levine, Martin Sasieta, Gautam
Satishchandran, Jon Sorce and Nico Valdes-Meller for illuminating discussions.
I also especially want to thank Netta Engelhardt for comments on an earlier
draft of the manuscript, Matt Headrick for extensive discussions leading to
Theorem~1, and Nico Valdes-Meller for many useful discussions.  This research is
supported by the John Templeton Foundation via the Black Hole Initiative. The
author is grateful to the MIT Center for Theoretical Physics for general
support.

\bibliographystyle{jhep}
\bibliography{all}

\end{document}